\documentstyle{article}
\newcommand{\mysection}[1]{\setcounter{equation}{0}\section{#1}}
\renewcommand{\theequation}{\thesection.\arabic{equation}}
\newcommand{\PSbox}[3]{\mbox{\rule{0in}{#3}\includegraphics{#1}\hspace{#2}}}

\begin{document}
\font\cmss=cmss10 \font\cmsss=cmss10 at 7pt
\hfill MIT-CTP-2588

\hfill hep-th/9611133

\hfill November, 1996
\vspace{1in}

\begin{center}
{\large {\bf \vspace{10pt}  Conformal Symmetry and the Chiral Anomaly}}\\%

\vspace{10pt}

{\sl Joshua Erlich}$^{a}$ and {\sl Daniel Z. Freedman}$^{b}$
\end{center}

\vspace{4pt}
$^{a}${\it Center for Theoretical Physics, Massachusetts Institute of
Technology, Cambridge MA 02139, USA}

$^{b}${\it Department of Mathematics and Center for Theoretical Physics,
Massachusetts Institute of Technology,}

{\it \ Cambridge MA 02139, USA}

\vspace{12pt}
\begin{center}
{\bf Abstract}
\end{center}
\vspace{4pt}

\noindent Two-loop contributions to the anomalous correlation function 
$\langle J_\mu(x) J_\nu(y) J_\rho(z)\rangle$ of three chiral 
currents are calculated by a method based on the conformal properties of
massless field theories. The method was previously applied
to virtual photon diagrams in quantum electrodynamics, and 
it is extended here to diagrams with scalars and chiral spinors in the abelian 
Higgs
model and in the $SU(3)\times SU(2)\times U(1)$ standard model.  In each case
there are nonvanishing contributions to the gauge current correlator from
self-energy insertions, vertex insertions and nonplanar diagrams, but their
sum exactly vanishes.  The two-loop contribution to the anomaly
therefore also vanishes, in agreement with the Adler-Bardeen theorem.  An
application of the method to the correlator $\langle R_\mu(x) R_\nu(y) K_\rho
(z)\rangle$ of the $R$ and Konishi axial currents in supersymmetric gauge 
theories which was reported in hep-th/9608125 is discussed here.  The net
two-loop contribution to this correlator also vanishes.
\vfill\eject
\setcounter{equation}{0}\vspace{0.2in}
\mysection{Introduction} 

   The chiral anomaly discovered long ago by Adler \cite{Adler} and Bell and 
Jackiw \cite{BellJackiw}
is a seminal concept of quantum field theory. The absence of radiative 
corrections to the one-loop anomaly is of central importance in applications
to neutral pion decay, to the structure of fermion families in the standard 
model, to mathematical contact between gauge theory and the Atiyah-Singer 
index theorem \cite{AtiyahSinger}, and many other questions. One might have 
thought that this
matter was settled by the early work of Adler and Bardeen \cite{AB} which 
involved regularizations of the theory, or by the general renormalization 
group argument \cite{Zee} for anomalies of global currents, or by BRS 
cohomology
arguments \cite{Piguet} for gauge current anomalies. Yet
there is much literature which
disputes the common wisdom \cite{DeRaad, Schwinger, AI, FMP}. Further a 
certain level of suspicion of 
general
theorems has proved to be healthy for theoretical physics, not necessarily
because proofs can be wrong, but because inappropriate assumptions can be made
in the hypotheses.  For example, the particular order of the operations of
regularization and computing the axial vector divergence which was used in
\cite{AB} can be questioned.

Thus explicit calculations of possible radiative corrections to the anomaly
in chiral gauge theories are illuminating.  A violation of the Adler-Bardeen
theorem in the standard model would be particularly significant because it
would call to question one of its most attractive features, namely that the 
one-loop
anomaly cancellation between quarks and leptons occurs so naturally and is
sufficient to make the theory consistent.  We therefore study two-loop
contributions to the gauge correlator $\langle J_\mu(x) J_\nu(y) J_\rho(z)
\rangle$ in the abelian Higgs model\footnote{Our investigation
was motivated by papers of Cheng and Li \cite{Cheng} in which 
a nonvanishing two-loop anomaly was obtained in this theory.  A subtle error
has recently been found \cite{Cheng-pc}, and there is now agreement on the vanishing of the
anomaly.}, which is a
simplified form of the standard model, and then extend the treatment to the
full glory of the $SU(3)\times SU(2)\times U(1)$ standard model, where there
are four
independent possibly anomalous correlators to be checked.  In all cases the
net sum of self-energy plus vertex insertions plus nonplanar diagrams vanishes.
So the full two-loop current correlators vanish, and their would be anomalous
divergences vanish, thus validating by explicit calculation the conventional
wisdom concerning radiative corrections to the chiral anomaly.

    In our method the current correlator is calculated directly in Euclidean
position space using a simplifying change of variables suggested by the
conformal properties of the correlator to perform the internal integrations.
Conformal symmetry also explains why the net two-loop correlator
$\langle J_\mu(x) J_\nu(y) J_\rho(z)\rangle$ vanishes, when one might have
expected only the vanishing of its divergence $\frac{\partial}{\partial x_\mu}
\,\langle J_\mu(x) J_\nu(y) J_\rho(z)\rangle$.  The two-loop correlator is
conformal covariant for massless internal lines, and one can show that for 
any conformal covariant
contribution the abnormal parity part of the third rank tensor correlator
vanishes if and only if its divergences vanish.

This method was previously used by Baker and Johnson \cite{BJ} to compute the 
two-loop vector and axial vector vertex functions in massless quantum 
electrodynamics. 
The ideas of the more comprehensive position space method of differential
renormalization \cite{Bible} also play a role, but the specific 
two-loop calculations
required to test the anomaly in this method do not require regularization or
cutoff. The basic ideas of the method are described in Section 2. The gauge
covariant derivative in the abelian Higgs model is 
\begin{equation} D_\mu\psi=(\partial_\mu+ig(\alpha L+\beta R)\,A_\mu)\psi 
\end{equation}
where $L$ and $R$ are chiral projectors.
We begin calculations in Section 3 at the point $\beta=-\alpha=1/2$, {\em i.e.}
 pure axial
coupling, because this is the point at which one can choose a gauge in which
the one-loop fermion vertex function and self-energy are finite.  This 
eliminates all subdivergences in the two-loop current correlator graphs. The
modifications required to handle all values of $\alpha,\beta$ are described in 
Section 
4.  Due to parity non-conservation there is no true finite gauge for the vertex
and self-energy functions, but we show that there is an effective finite
gauge in which the two-loop vertex and self-energy insertion contributions to
$\langle J_\mu(z)J_\nu(x)J_\rho(y)\rangle$have no
subdivergences.  In Section 5 the method
is extended to the $SU(3)\times SU(2)\times U(1)$ standard model. We assume 
the usual 
couplings for which the one-loop anomalies cancel. There are then
no genuinely new graphs to compute, but the effective finite gauge mechanism
is more complicated than before. In supersymmetric gauge theories there are
two formally conserved axial currents: the $R$-charge current $R_\mu(x)$ and 
the
Konishi current $K_\mu(x)$. The correlator $\langle R_\mu(x)  R_\nu(y) K_\rho
(z)\rangle$ was 
calculated by the present
methodology as part of a recent study \cite{SCFT} of the OPE's of the superconformal
algebra. Details were not discussed in \cite{SCFT}, and they are briefly presented
in Section 6 below.  

\setcounter{section}{1}
\mysection{The Method}

Although conformal symmetry is concretely used in our work largely to motivate a
change of variables which simplifies the required two-loop Feynman integrals,
we believe that it is useful to explain the method from a more fundamental  
standpoint.  It is well known (see, for example, \cite{Osborn1}) that the
conformal group of Euclidean field theory is $O(5,1)$, and that all
transformations which are continuously connected to the identity are
obtained by combining rotations and translations with the basic conformal
inversion \begin{eqnarray}
x_\mu&=&\frac{x'_\mu}{x'^2} \nonumber \\
\frac{\partial x_\mu}{\partial x'_\nu}&=&x^2 \left(\delta_{\mu\nu}-\frac{2x_\mu
x_\nu}{x^2}\right)\equiv x^2\,J_{\mu\nu}(x) \, .\end{eqnarray}
The Jacobian tensor $J_{\mu\nu}(x)=J_{\mu\nu}(x')$, which is an improper
orthogonal matrix, will be very useful for us.  Because Det $J=-1$, the
inversion is a discrete operation \cite{Osborn1}, similar to parity, and not
an element of the continuous component of $O(5,1)$ which contains the identity.

The Euclidean action of the massless $U(1)$ Higgs model is
\begin{eqnarray}
&S=\int d^4x\,\left[\frac{1}{4}F_{\mu\nu}^2+D_\mu\overline{\phi}\,D_\mu\phi+
\overline{\psi}\gamma_\mu D_\mu\psi-f\,\overline{\psi}(L\phi+R\overline{\phi})
\psi-\frac{\lambda}{4}(\overline{\phi}\phi)^2\right]& \nonumber \\ \vspace{2pt}
&D_\mu\phi=(\partial_\mu+igA_\mu)\phi\hspace{10mm}D_\mu\psi=\left(\partial
_\mu+igA_\mu\,\left(\alpha L+\beta R\right)\right)\psi& \label{eq:2.2} \\
&\beta-\alpha=1\hspace{10mm}\gamma_5=\gamma_1\gamma_2\gamma_3\gamma_4\hspace
{10mm}L=\frac{1}{2}(1-\gamma_5)\hspace{10mm}R=\frac{1}{2}(1+\gamma_5)&
\nonumber \end{eqnarray}
It is invariant under conformal transformations in the continuous component of
$O(5,1)$, but not necessarily under inversion since that question is related
\cite{Schreier} to invariance under discrete symmetries.  For the special
choice $\beta=-\alpha=1/2$, where we have a parity conserving theory with pure
axial gauge coupling, invariance holds under the transformations
\begin{eqnarray}
\phi(x)&\rightarrow&\phi'(x)=x'^2\overline{\phi}(x') \nonumber \\
\psi(x)&\rightarrow&\psi'(x)=x'^2\gamma_5/\hspace{-2mm}x'\psi(x') \\
\overline{\psi}(x)&\rightarrow&\overline{\psi'}(x)=x'^2\,\overline{\psi}/\hspace{-2mm}x'
\gamma_5 \\
A_\mu(x)&\rightarrow& A_\mu'(x)=-x'^2\, J_{\mu\nu}(x')A_\nu(x')\, ,\nonumber
\end{eqnarray}
as can be verified with diligence and the help of the relations 
\begin{equation}
d^4x=\frac{d^4x'}{x'^8}\hspace{1cm}/\hspace{-2mm}x'\gamma_\mu/\hspace{-2mm}x'
=-x'^2\, J_{\mu\nu}(x')\gamma_\nu \, .\label{eq:2.4}\end{equation}
Inversion invariance does not hold in the general
chiral theory, and it is not required for our application.

It is important that correlation functions are constructed from Feynman
rules in which the vertex factors and propagators have simple inversion
properties.  In particular the scalar and spinor propagators transform as
\begin{eqnarray}
&{\displaystyle \Delta(x-y)=\frac{1}{4\pi^2}\frac{1}{(x-y)^2}=\frac{1}{4\pi^2}\frac{x'^2y'^2}
{(x'-y')^2}}\, .&\nonumber \\
&{\displaystyle S(x-y)=-/\hspace{-2mm}\partial\Delta(x-y)=\frac{1}{2\pi^2}\frac{/\hspace
{-2mm}x-/\hspace{-2mm}y}{(x-y)^2}=-\frac{1}{2\pi^2}x'^2y'^2/\hspace{-2mm}x'
\frac{(/\hspace{-2mm}x'-/\hspace{-2mm}y')}{(x'-y')^4}/\hspace{-2mm}y'}\, .&
\label{eq:2.5}\end{eqnarray}
The gauge field propagator is another story \cite{BJ}. In the usual family of
covariant gauges one has
\begin{equation}
\Delta_{\mu\nu}=\frac{1}{4\pi^2}\left[\frac{\delta_{\mu\nu}}{(x-y)^2}-\frac{1}
{2}\Gamma\, \frac{J_{\mu\nu}(x-y)}{(x-y)^2}\right] \, 
\label{eq:2.6}\end{equation}
where $\Gamma=0$ is the Feynman gauge and $\Gamma=1$ is the Landau gauge. Only the
second term transforms as expected under inversion, since \begin{equation}
J_{\mu\nu}(x-y)=J_{\mu\rho}(x')\,J_{\rho\sigma}(x'-y')\,J_{\sigma\nu}(y')
\, .\label{eq:2.7}\end{equation}
The full propagator transforms properly only after a gauge transformation is
performed \cite{BJ}, and this complicates applications to amplitudes with virtual
photons.

It is well known that conformal symmetry restricts the tensorial form of 
two and three-point correlation functions and frequently determines these
tensors uniquely up to a constant multiple. (For recent discussions, see
\cite{Osborn1,Osborn2}). Inversion symmetry is sufficient to determine these
restrictions, and the inversion property of a vector current of dimension 3 is
\begin{equation}
J_{\mu}(x)\rightarrow J_\mu'(x')=x'^6J_{\mu\nu}(x')\,J_\nu(x') \, .
\label{eq:2.8}\end{equation}
We are primarily interested in the abnormal parity part of
the correlator $\langle J_\mu(x)J_\nu(y)J_\rho(z)\rangle$ of three conserved currents, and it is
known that there is \cite{Schreier} a unique conserved rank 3 tensor function 
with the inversion property required by (\ref{eq:2.8}). The specific form is given,
up to a multiplicative constant, by the lowest order massless fermion axial 
triangle amplitude (Fig. 1a)
\begin{eqnarray}
A_{\mu\nu\rho}(z,x,y)&=&{\scriptstyle (-)}{\rm Tr}\,\gamma_\mu\gamma_5\,S(z-y)
\,\gamma_\rho\,S(y-x)\,\gamma_\nu\,S(x-z) \nonumber \\
 &=&\frac{1}{(2\pi^2)^3}\,{\rm Tr}\left\{\gamma_5\gamma_\mu\frac{/\hspace{-2mm}z
-/\hspace{-2mm}y}{(z-y)^4}\,\gamma_\rho\,\frac{/\hspace{-2mm}y-/\hspace{-2mm}x}
{(y-x)^4}\,\gamma_\nu\,\frac{/\hspace{-2mm}x-/\hspace{-2mm}z}{(x-z)^4}\right\}
\, ,\label{eq:2.9}
\end{eqnarray}
in which the (-) is the usual factor for a closed fermion loop. The conformal
properties can be readily verified using (\ref{eq:2.4}-\ref{eq:2.5}).

For separated points this function obeys all desiderata. It is fully Bose 
symmetric and conserved on all three indices. The expected anomaly is a local
violation of the conservation Ward identities which arises because the 
differentiation
of singular functions is involved. There are several ways \cite{Bible,
Stony,Sonoda} to obtain the anomaly from this $x$-space viewpoint. One way 
\cite{Bible},
which we now summarize, is
to recognize that the amplitude (\ref{eq:2.9}) is too singular at short distance to
have a well defined Fourier transform. One then regulates which entails the
introduction of several independent mass scales, but the regulated form after 
the gamma matrix trace depends only on 
the ratio of these scales. The regulated amplitude is well defined, and
one can check the Ward identities, which take the expected form
\begin{equation}
\frac{\partial}{\partial z^\mu}\,A_{\mu\nu\rho}(z,x,y)=a_z\,\varepsilon_{\nu\rho
\lambda\sigma}\,\frac{\partial}{\partial x_\lambda}\frac{\partial}{\partial y_
\sigma}\,\delta(x-z)\delta(y-z) \, ,
\end{equation}
with similar expressions for the divergences with respect to $x_\nu$ and 
$y_\rho$.
The anomaly coefficients $a_z, a_x, a_y$ depend on the ratio of mass scales,
and there is no choice of scales which makes all coefficients vanish.
Specifically the sum $a_x +a_y +a_z = -1/4\pi^2$ is independent of the scales. There
is a choice of scales which makes $a_x=a_y=a_z=-1/12\pi^2$ which is the Bose
symmetric choice relevant for the gauge current correlation function in the
$U(1)$ Higgs model (\ref{eq:2.2}), and another choice to make $a_x=a_y=0$ which is 
appropriate for the correlator of one axial and two vector currents.

A tenet of the space-time approach to renormalization is
that the intrinsic ambiguity of a primitively divergent amplitude is an
ultra-local distribution consistent with dimension and symmetry requirements.
This corresponds to the p-space ambiguity of polynomials in external momenta.
In this light the ambiguous part of the tensor amplitude (\ref{eq:2.9}) is 
\begin{equation}
\Delta A_{\mu\nu\rho}=\varepsilon_{\mu\nu\rho\sigma}\left[b_1\left(\frac
{\partial}{\partial x_\sigma}-\frac{\partial}{\partial y_\sigma}\right)+b_2\left(
\frac{\partial}{\partial y_\sigma}-\frac{\partial}{\partial z_\sigma}\right)
\right]\delta(x-z)\delta(y-z)\, ,
\end{equation}
where $b_1$ and $b_2$ are arbitrary constants. The mass scale dependence of the
regulated amplitude is exactly of this form, and its Fourier transform is just
the shift ambiguity due to choice of loop momenta in the traditional approach
to the anomaly \cite{Jackiw}. The nugget of this discussion of the space-time
approach to the lowest order axial anomaly is that the well defined amplitude 
(\ref{eq:2.9}) for separated points determines the fact that there is an anomaly of
specific strength. The choice of regularization or calculational procedure
for the Fourier transform is just a redistribution of the anomaly between the
three parameters $a_x, a_y, a_z$, which does not affect their sum.

We now return to the discussion of conformal symmetry and its role in the
elucidation of possible radiative corrections to the anomaly. One may question
this role because of the common lore that the introduction of a scale required
to handle the divergences of perturbation theory spoils expected conformal 
properties. In general this is true, but the two-loop anomaly diagrams of
Higgs models, which are drawn in Fig. 1, are exceptional. Any primitively 
divergent amplitude is exceptional when studied in $x$-space for separated
points, since the internal integrals converge without regularization. The
nonplanar diagrams of Fig. 1h are primitives. Of course there are many other 
diagrams which contain sub-divergent vertex and self-energy corrections, and
these require a regularization scale. However for the specific choice $\beta=
-\alpha=1/2$ which corresponds to pure axial coupling for the fermion, it is
quite easy to see that there is a unique choice of gauge-fixing parameter
$\Gamma$ which makes the one-loop self energy finite. Since the vertex and 
self-energy corrections are related by a Ward identity, each vertex correction 
is also finite in the same gauge; specifically, the sum of the three contributing
Feynman diagrams is ultraviolet finite. In this ``finite gauge'' the integrals
in the sum of three vertex insertion diagrams (Fig.~1b,c,d) at each corner of 
the two-loop
triangle converge. The same statement holds for the integrals in the sum of
the two self-energy insertion diagrams (Fig.~1e,f) on each leg of the 
triangle.

The photon propagator (\ref{eq:2.6}) is not conformal covariant, and we will discuss
this complication in the next section. We will show there that the diagrams
for which this difficulty occurs are already covered by previous work 
\cite{BJ, ABWY}.
In the remaining diagrams the photon propagator may be replaced by the 
inversion covariant second term of (\ref{eq:2.6}) and a finite gauge can be chosen. We
then have the situation that each two-loop Feynman diagram we need to compute
is constructed with inversion covariant propagators and vertices, and the 
sums of the self-energy diagrams on each leg and vertex diagrams at each 
corner are convergent. Then each of the three nonplanar diagrams and the summed
self-energy or vertex insertion diagrams at each leg or corner of the 
triange is a conformal covariant contribution to the current correlation 
function. Each of these amplitudes must be a multiple of the unique conformal 
tensor $A_{\mu\nu\rho}$ of (\ref{eq:2.9}), and we will show that the sum of the separate
contributions to the net two-loop correlator vanishes. Further, we will
show that the conformal inversion can be used as a transformation of the 
integration
variables which makes the calculation of the eight-dimensional integrals easy
and also gives an explicit verification of the conformal properties we have
discussed above.    

\setcounter{section}{2}
\mysection{The $U(1)$ model for $\beta=-\alpha= 1/2$}

Euclidean correlation functions for the theory are constructed using
the propagators of (\ref{eq:2.5}) and (\ref{eq:2.6}), the vertex rules which
can be read from the action (\ref{eq:2.2}), and the instruction to integrate 
$\int d^4u$ over each internal vertex of a diagram.

To illustrate the way conformal symmetry is used in our work, 
we first study the non-planar graph of Fig. 1h.  
\begin{figure}\begin{center}
\PSbox{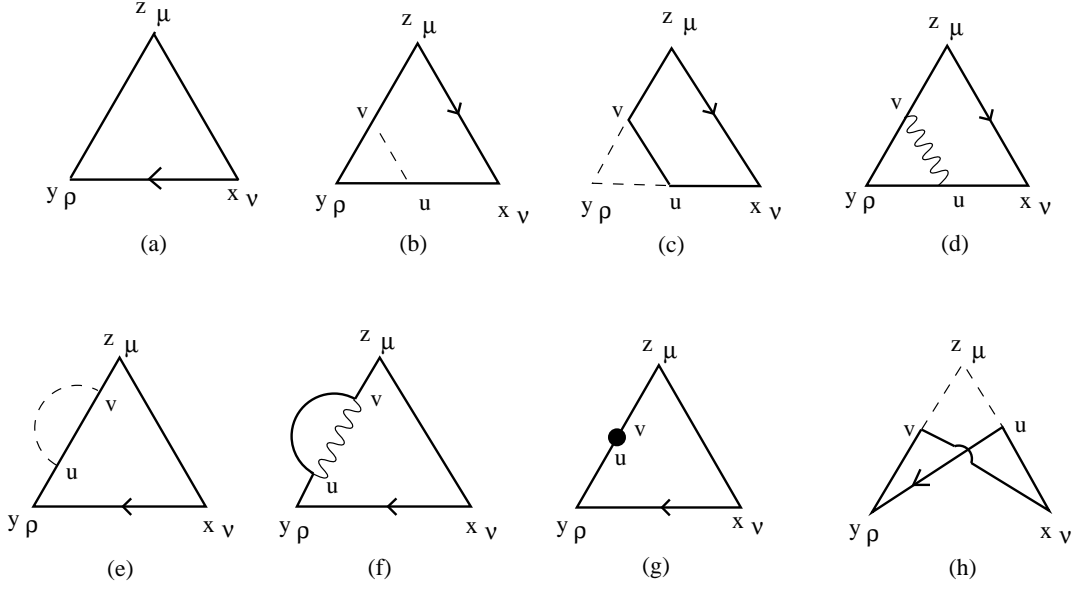}{5.3in}{2.7in} \end{center}
\caption{One and two-loop contributions to the anomaly in the abelian Higgs theory.  
There is a chiral current at each
corner of the triangle.  Solid lines are fermions, dashed lines scalars and
wavy lines gauge fields.  The filled dot in (g) is a local 
self-energy renormalization.  Not shown are the same 
two-loop diagrams rotated
$\pm 120^\circ$ and diagrams with the opposite direction of fermion flow for
all but the nonplanar diagram (h), and it is understood that in (b), (c), (e)
and (h) both directions of Higgs propagation are included.  
The coordinate and
Lorentz indices correspond to the integrals in Section 3, and we refer
to (b)-(h) with and without these indices (by their general topology)
 in the text.}
\label{fig:1}
\end{figure}
Its amplitude
is conformal covariant since no issues of subdivergences and gauge
choice arise.  The idea is to use the inversion, 
$u_{\alpha} = u'_{\alpha}/ u'^{2}$ and 
$v_{\alpha} = v'_{\alpha}/ v'^{2}$, as a change of variable in
the internal integrals.  In order to use the simple conformal
properties of the propagators (\ref{eq:2.5}) we must also refer the
external points to their inverted images, e.g.\ 
$x_{\mu} = x'_\mu/ x'^{2}$.  
If this is done for a generic configuration of $x,
y, z$, there is nothing to be gained because the same integral is
obtained in the $u', v'$ variables.   However if
we use translation symmetry to place one point at $0$, say $z =
0$, it then turns out that the propagators attached to that point
drop out of the integral, essentially because the inverted point
is at $\infty$, and the integrals simplify.

After summing over both directions of Higgs field propagation and
elementary manipulation of chiral factors $L$ and $R$, the
amplitude for the graph (Fig. 1g) can be written as
\begin{eqnarray}
  \lefteqn{N_{\mu\nu\rho} \left(0, x, y \right)} 
\nonumber    \\
  {} &=& \frac{i g^{3} f^{2}} {512\pi^{12}}
     \int d^{4} u\,d^4 v
     \left( \frac{v_{\mu}}{u^{2}v^4}-\frac{u_\mu}{u^4v^2} \right)
     {\rm Tr}\,\gamma_{5} \gamma_{\rho} 
     \frac{/\hspace{-2mm}y -/\hspace{-2mm}u} {\left(y - u\right)^{4} }
     \frac{/\hspace{-2mm}u - /\hspace{-2mm}x}{\left(u - x \right)^{4}}
     \gamma_\nu 
     \frac{/\hspace{-2mm}x - /\hspace{-2mm}v}{\left(x - v \right)^{4}} 
     \frac{/\hspace{-2mm}v - /\hspace{-2mm}y}{\left(v - y \right)^{4}}\, ,
\label{eq:3.1} \end{eqnarray}
The integration variables $u, v$ each appear in three
denominators.  This is not necessarily fatal, and indeed the
$u$ and $v$ integrals can be evaluated in closed form
using Feynman parameters \cite{Cheng-pc}.  However, we will see
that the conformal inversion leads to simpler integrals.  The
change of variables outlined in the previous paragraph can be
made with the help of (\ref{eq:2.4}--\ref{eq:2.5}) and the Higgs current
transformation
\begin{equation}
  \label{eq:3.2}
  \frac{v_{\mu}}{u^{2} v^{4}} 
  - \frac{{u}_{\mu}} {u^{4} v^{2}} 
  = u'^{2} v'^{2}
  \left(v'_{\mu} - u'_{\mu} \right)\, .
\end{equation}


Spinor propagator ``side factors'' {\em e.g.}\ $/\hspace{-2mm}x', /\hspace
{-2mm}u'$,
etc.\ all collapse within the trace, and the Jacobian
$\left(u'v'\right)^{-8}$ cancels with factors in the
numerator giving the result
\begin{eqnarray}
  N_{\mu\nu\rho} \left(0,x,y\right) 
  &=&  \frac{i g^{3} f^{2}}{512\pi^{12}}\, 
       x'^{6} y'^{6}\, J_{\nu\nu'} (x') J_{\rho\rho '} (y')\,
       \tilde{N}_{\mu \nu' \rho'} 
       \nonumber \\
  \tilde{N}_{\mu\nu'\rho'} 
  &=& \int d^4u'\,d^4v'\,(v'_\mu-u'_\mu)\,{\rm Tr}\left[\gamma_5\gamma_{\rho'}
\,\frac{/\hspace{-2mm}y'-/\hspace{-2mm}u'}{(y'-u')^4}\frac{/\hspace{-2mm}u'
-/\hspace{-2mm}x'}{(u'-x')^4}\,\gamma_{\nu'}\,\frac{/\hspace{-2mm}x'-
/\hspace{-2mm}v'}{(x'-v')^4}\frac{/\hspace{-2mm}v'-/\hspace{-2mm}y'}{(v'-y')
^4}\right]\, .
  \label{eq:3.3}  
\end{eqnarray}
We see the expected transformation factors for the currents at
$x$ and $y$ times an integral in which $u'$ and $v'$ each appear
in only two denominators.  Such convergent tensorial convolution
integrals can be done by several methods.  We have used
Gegenbauer polynomial methods \cite{Rosner}, and the results are tabulated in
Appendix A.  When these results are used and substituted
within the trace, one finds the final amplitude
\begin{equation}
  \label{eq:3.4}
  N_{\mu \nu \rho} \left(0, x, y \right) = -\frac{i g^3 f^2}
  {512\pi^8}\, 
  x'^{6}y'^{6} \,J_{\nu\nu'}(x')\,J_{\rho
      \rho'}\left({y'}\right)\,
    \frac {{\rm Tr}\, \gamma_{5} \gamma_{\mu} \gamma_{\rho'} \gamma_{\nu'}\,
      \left(/\hspace{-2mm}x' - /\hspace{-2mm}y' \right)} 
{\left( x' - y' \right)^4}\, .
\end{equation}

The result above may be compared with the amplitude of the
one-loop triangle graph (Fig. 1a) (with one direction of charge flow).
\begin{equation}
  \label{eq:3.5}
B_{\mu\nu\rho}(z,x,y)=\frac{ig^3}{8}A_{\mu\nu\rho}(z,x,y)\, ,
\end{equation}
where $A_{\mu\nu\rho}$ is given in (\ref{eq:2.9}).
At $z=0$, and referred to inverted points $x',y'$, this reads
\begin{equation}
B_{\mu\nu\rho}(0,x,y)=\frac{ig^3}{8(8\pi^6)}\,x'^6y'^6\,J_{\nu\nu'}(x')
J_{\rho\rho'}(y')\,\frac{{\rm Tr}\,\gamma_5\gamma_\mu\gamma_{\rho'}\gamma_{\nu'}
\,(/\hspace{-2mm}x'-/\hspace{-2mm}y')}{(x'-y')^4}\, .
\label{eq:3.6}
\end{equation}
One may now observe that the non-planar amplitude is just a numerical
multiple of the unique conformal tensor (\ref{eq:2.9}) as discussed in
Section~2.  The result may be written  \begin{equation}
N_{\mu\nu\rho}(0,x,y)=-\frac{f^2}{8\pi^2}\,B_{\mu\nu\rho}(0,x,y)\, .
\label{eq:3.7} \end{equation}

The non-planar graphs with scalar vertices at $x$ and $y$ must give the 
same result by triangular symmetry.  However, our method of evaluation of the
amplitude has singled out the point $z=0$.  Therefore a check on
the result can be determined by applying the inversion to the
amplitudes for the $\pm 120^\circ$ rotated diagrams with $z=0$.  The integral 
in inverted
variables involves a different set of convolution integrals, and we
have checked that it gives the same result (\ref{eq:3.7}).

We now discuss, following \cite{BJ}, the finite gauge mechanism
for the one-loop self-energy and vertex corrections which are
ingredients of our study of the two-loop anomaly.  After a
little algebra the sum of the Higgs and photon self-energy graphs
can be written as
\begin{equation}
  \label{eq:3.9}
  \Sigma(v-u) = \frac{1}{8 \pi^{4}} 
  \left[ f^{2} + \frac{1}{2} g^{2} 
      \left( 1-\Gamma \right) 
  \right]
  \frac{/\hspace{-2mm}v -/\hspace{-2mm}u}{(v-u)^6}
  + a\, /\hspace{-2mm}\partial\, \delta^4 
      \left(v-u\right)\, .
\end{equation}
The first term is the separated point part of the amplitude which
is completely determined by the Feynman rules.  It is a
singular function of $v-u$ whose Fourier transform is linearly
divergent.  By choosing the gauge $\Gamma = 1 + {2f^{2}}/g^{2}$,
the amplitude is made finite.  It vanishes for separated points,
but there is a possible local term, the second term in (\ref{eq:3.9}), which
is left ambiguous by the Feynman rules.  The finite constant $a$
will be determined by the Ward identity.  

The amplitudes of the three vertex subgraphs in the diagrams (Fig.~1b,c,d) are
\begin{eqnarray}
  \label{eq:3.10}
 &{\displaystyle V^{(1)}_\rho (y,v,u) = 
  \frac{-igf^{2}}{32\pi^{6}}\,\gamma_5\, 
  \frac{1}{\left(v-u\right)^{2}} 
  \frac{/\hspace{-2mm}v -/\hspace{-2mm}y}{\left(v-y\right)^{4}}\,
  \gamma_{\rho} 
  \frac{\left(/\hspace{-2mm}y -/\hspace{-2mm}u\right)}{\left(y-u\right)^{4}}}
\, ,&
\\
 &{\displaystyle
V_\rho^{(2)}(y,v,u)=-\frac{igf^2}{32\pi^6}\,\gamma_5\,\frac{1}{(v-y)^2}\,\frac{\stackrel
{\longleftrightarrow}{\partial}}{\partial y_\rho}\,\frac{1}{(y-u)^2}\,\frac{
/\hspace{-2mm}v-/\hspace{-2mm}u}{(v-u)^4}}\, .& \label{eq:3.11} \\
 &{\displaystyle V_\rho^{(3)}(y,v,u)=\frac{ig^3}{128\pi^6}\,\gamma_5\gamma_\alpha\,\frac{
/\hspace{-2mm}v-/\hspace{-2mm}y}{(v-y)^4}\,\gamma_\mu\,\frac{/\hspace{-2mm}y-
/\hspace{-2mm}u}{(y-u)^4}\,\gamma_\beta\left[\frac{\delta_{\alpha\beta}\left(
1-\frac{1}{2}\Gamma\right)}{(u-v)^2}+\frac{\Gamma\,(u-v)_\alpha(u-v)_
\beta}{(u-v)^4}\right]}\, .& \label{eq:3.12} \end{eqnarray}
Each contribution has a logarithmic divergent Fourier transform, and we
consider the Fourier transform at zero fermion momentum to
study finiteness properties; that is, we consider the integrals
$\int d^4u\,d^4v\,V_\rho(y,v,u)$.  Let us examine first the single integral
$\int d^4v\,V_\rho(y,v,u)$ which can be simplified by taking the point $y=0$.
We will discuss these integrals in some detail because the same integrals
will occur in the vertex insertion diagrams of the two-loop current correlator.

We see that the required integrals are again convergent convolution integrals.
Using the tabulation in Appendix A, we obtain \begin{eqnarray}
\int d^4v\,V_\rho^{(1)}(0,v,u)&=&\frac{igf^2}{32\pi^4}\,\frac{\gamma_5\,
/\hspace{-2mm}u\,\gamma_\rho\,/\hspace{-2mm}u}{u^6} \nonumber \\
 &=&-\frac{igf^2}{32\pi^4}\,\gamma_5\gamma_\sigma\,\left(\frac{\delta_{\sigma
\rho}}{u^4}-\frac{2u_\sigma u_\rho}{u^6} \, .\label{eq:3.13}\right) \\
\int d^4v\,V_\rho^{(2)}(0,v,u)&=&\frac{igf^2}{32\pi^4}\,\frac{\gamma_5\gamma_
\rho}{u^4} \, .\label{eq:3.14} \end{eqnarray}
\begin{eqnarray}
\int d^4v V_\rho^{(3)}(0,v,u)&=&\frac{ig^3}{128\pi^4}\,\frac{\gamma_5\gamma_\alpha
\gamma_\sigma\gamma_\rho /\hspace{-2mm}u\gamma_\beta}{u^4}\left[\left(1-\frac{1}
{4}\Gamma\right)\frac{\delta_{\alpha\beta}\,u_\sigma}{u^2}-\frac{1}{4}\Gamma
\left(\frac{\delta_{\alpha\sigma}u_\beta+\delta_{\beta\sigma}\,u_\alpha}{u^2}
-2\frac{u_\alpha u_\beta u_\sigma}{u^4}\right)\right] \nonumber \\
 &=&\frac{ig^3}{128\pi^4}\,\frac{\gamma_5\gamma_\alpha}{u^6}\left[\left(u^2
\delta_{\alpha\rho}-4u_{\alpha\rho}\right)+(1-\Gamma)\delta_{\alpha\mu}\,u^2\right]\, .
\label{eq:3.15} \end{eqnarray}

Consider next the second integration $\int d^4u\int d^4v\, V_\rho(0,v,u)$
which gives the zero momentum vertex function.  The $\delta_{\alpha\rho}/u^4$
terms give the expected logarithmic divergence, but the integrals over the
traceless tensor $(u^2\delta_{\alpha\rho}-4u_\alpha u_\rho)/u^6$ converge by
symmetric integration.  One sees that the sum of the divergent contributions
from (\ref{eq:3.13}), (\ref{eq:3.14}) and (\ref{eq:3.15}) is proportional
to \begin{equation}
2f^2-4f^2-g^2(1-\Gamma) \, ,\label{eq:3.16} \end{equation}
and therefore vanishes in the same gauge that makes the self-energy finite.
Henceforth we will use this gauge.

The Ward identity for the theory may be derived by standard functional methods
or obtained directly from the vertex amplitudes (\ref{eq:3.10}-\ref{eq:3.12})
using the relation \begin{equation}
\Box \frac{1}{(y-u)^2}=-4\pi^2\,\delta^4(y-u) \, .\label{eq:3.17}\end{equation}
The result is \begin{equation}
\frac{\partial}{\partial y_\rho}\,V_\rho(y,v,u)=-i\frac{1}{2}g\gamma_5\,(
\delta^4(y-v)-\delta^4(y-u))\,\Sigma(v-u) \, .\label{eq:3.18} \end{equation}
We wish to determine the constant $a$ in the self-energy (\ref{eq:3.9}).  This
would appear in (\ref{eq:3.18}) as the coefficient of a very singular 
distribution, so we integrate with respect to the smooth test function $1$ and use the
integrated Ward identity \begin{equation}
\frac{\partial}{\partial y_\rho}\int d^4v\,V_\rho(y,v,0)=-i\frac{1}{2}g
\gamma_5\,\Sigma(y) \, .\label{eq:3.19} \end{equation}

In the finite gauge the only contributions to the integral in (\ref{eq:3.19})
come from the traceless tensor structures in (\ref{eq:3.13}) and (\ref
{eq:3.15}), and we have \begin{equation}
\int d^4v\,V_\rho(y,v,0)=\frac{ig}{16\pi^4}\,(f^2-\frac{1}{2}g^2)\,\gamma_5
\gamma_\sigma\left(\frac{y_\sigma y_\rho}{y^6}-\frac{\delta_{\sigma\rho}}{4y^4}
\right) \, .\label{eq:3.20} \end{equation}
The singular tensor can be expressed as the traceless form \begin{equation}
\frac{y_\sigma y_\rho}{y^6}-\frac{\delta_{\sigma\rho}}{4y^4}=\frac{1}{8}\left(
\partial_\sigma\partial_\rho-\frac{1}{4}\,\delta_{\sigma\rho}
\,\Box\right)\,\frac{1}{y^2}\, , \label{eq:3.21} \end{equation}
which is a well defined distribution, and its divergence is easily obtained:
\begin{equation}
\frac{\partial}{\partial y_\rho}\,\left(\frac{y_\sigma y_\rho}{y^6}-\frac{\delta_{\sigma\rho}}{4y^4}\right)=-\frac{3\pi^2}{8}\,\partial_\sigma\,\delta^4(y)
\, .\label{eq:3.21.5} \end{equation}
Combining (\ref{eq:3.19}-\ref{eq:3.21.5}) one finds the self-energy in the finite
gauge \begin{equation}
\Sigma(y)=\frac{3}{64\pi^2}(f^2-\frac{1}{2}g^2)\,/\hspace{-2mm}\partial\,
\delta^4(y) \, .\label{eq:3.22} \end{equation}

It is this result for $\Sigma(v-u)$ which is to be used to evaluate the 
self-energy insertion contributions (Fig.~1g) to the two-loop anomalous current
correlation function.  Because (\ref{eq:3.22}) is purely local, the integral
$\int d^4u\,d^4v$ required for the graph of Fig.~1g is trivial.  Specifically
the $\delta^4(v-u)$ in (\ref{eq:3.22}) can be used to do the $u$-integral, and
$/\hspace{-2mm}\partial_v$ acts on the resulting propagator giving a second
$\delta$-function which can be used to do the $v$-integral.  The result is that
the sum of the self-energy insertion graphs (Fig.~1e,f,g) is a multiple of the one-loop
amplitude \begin{equation}
\Sigma_{\mu\nu\rho}(z,x,y)=\frac{3}{64\pi^2}\left(f^2-\frac{1}{2}g^2\right)
B_{\mu\nu\rho}(z,x,y) \label{eq:3.23} \, .\end{equation}

Let us now discuss the diagrams which remain to be evaluated.  The diagrams
(Fig. 2a) with three virtual boson lines vanish trivially because the fermion
trace vanishes.  Next come the vertex insertion diagrams.
It is convenient to view each virtual photon diagram as the sum of two graphs,
one with the photon propagator in the Landau gauge $\Gamma=1$, and the second
with inversion covariant pure gauge propagator
\begin{equation}
\tilde{\Delta}_{\mu\nu}(u-v)=-\frac{1}{8\pi^2}\frac{\gamma\,f^2}{g^2}\frac{1}{(u-v)
^2}\,J_{\mu\nu}(u-v)\, ,\hspace{0.2in}\gamma=2\, .
 \label{eq:3.24}\end{equation}
The Landau gauge graphs give order $g^5$ contributions to the two-loop
$\langle J_\mu J_\nu J_\rho\rangle$ correlator, and the remainder gives an
order $g^3f^2$ contribution.  We now discuss these separately.
\begin{figure}\begin{center}
\PSbox{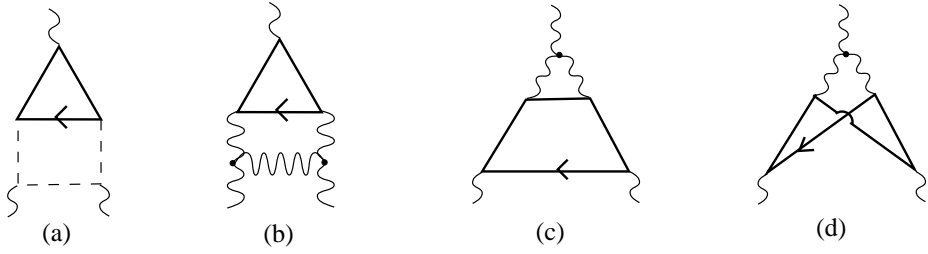}{5in}{1.5in}
\end{center}
\caption{Vanishing contributions to the three-current correlators.  Diagrams
(b), (c) and (d) are only present in correlators with nonabelian gauge
currents, and are discussed in Section 5.2.}
\end{figure}
The Landau gauge propagator is conformal covariant only after regauging by
adding a gradient term given in \cite{BJ}.  So the sum of all virtual photon 
diagrams is conformal covariant, but individual virtual photon vertex 
insertion diagrams (Fig.~1d) are not, and are not simplified by the simple
inversion discussed at the beginning of this section.  Nevertheless the photon
vertex contributions to the two-loop $\langle A_\mu V_\nu V_\rho\rangle$
correlator in quantum electrodynamics were calculated by related but more
complicated techniques in \cite{BJ}.  The net contribution to the correlator of
vertex and self-energy insertions was found to vanish there, thus verifying
the Adler-Bardeen theorem through two-loop order in QED.  It is quite easy
to see that, after clearing $\gamma_5$ factors, all Landau gauge virtual photon
diagrams of Fig.~1d in the axial coupled abelian Higgs model are a uniform
factor of $1/8$ times the same graphs in QED.  So it is fortunate that the
work of \cite{BJ} can be taken over to our case with the immediate result that the
net contribution of Landau gauge virtual photon graphs to the gauge current
correlator $\langle J_\mu J_\nu J_\rho\rangle$ vanishes.
Specifically the sum of the Landau gauge vertex insertions and the order $g^5$
part of the self-energy insertions (second term in \ref{eq:3.23}) vanishes.

We now study the order $g^3f^2$ vertex insertion contributions to the 
three-point current correlator.  These include virtual Higgs diagrams plus
a virtual photon diagram with the propagator (\ref{eq:3.24}).
Each of these graphs has a conformal covariant integrand, so the inversion
technique can be applied.
The amplitude for the diagram shown in Fig.~1b is, with $z=0$, \begin{equation}
V_{\mu\nu\rho}^{(1)}(0,x,y)=\frac{ig^3f^2}{512\pi^{12}}\int\frac{d^4u\,d^4v}
{(u-v)^2}\,{\rm Tr}\,\left[\gamma_5\,\frac{/\hspace{-2mm}v-/\hspace{-2mm}y}
{(v-y)^4}\,\gamma_\rho\,\frac{/\hspace{-2mm}y-/\hspace{-2mm}u}{(y-u)^4}\frac{
/\hspace{-2mm}u-/\hspace{-2mm}x}{(u-x)^4}\,\gamma_\nu\,\frac{/\hspace{-2mm}x}
{x^4}\,\gamma_\mu\,\frac{/\hspace{-2mm}v}{v^4}\right]\, . \label{eq:3.25}
\end{equation}
The inversion may be performed; Jacobian factors again cancel and the trace
simplifies giving \begin{eqnarray}
 & &V_{\mu\nu\rho}^{(1)}=\frac{ig^3f^2}{512\pi^{12}}x'^6y'^6\,J_{\nu\nu'}(x')
J_{\rho\rho'}(y')\tilde{V}_{
\mu\nu'\rho'}^{(1)} \nonumber \\ \vspace{4pt}
\lefteqn{\tilde{V}_{\mu\nu'\rho'}^{(1)}=
\int \frac{d^4u'\,d^4v'}{(u'-v')^2}\,{\rm Tr}\,\left[
\gamma_5\,\frac{/\hspace{-2mm}v'-/\hspace{-2mm}y'}{(v'-y')^4}\,\gamma_{\rho'}\,
\frac{/\hspace{-2mm}u'-/\hspace{-2mm}x'}{(u'-x')^4}\,\gamma_{\nu'}\gamma_\mu
\right]\, .} \label{eq:3.26} \end{eqnarray}
In both (\ref{eq:3.25}) and (\ref{eq:3.26}) the first three factors in the
integrals are exactly those of the vertex function $V_\rho^{(1)}(y,v,u)$.  The
major difference between (\ref{eq:3.25}) and (\ref{eq:3.26}) is that the
$/\hspace{-2mm}x/x^4$ and $/\hspace{-2mm}v/v^4$ propagators have disappeared.
The variable $v'$ now appears in two denominators, and the $v'$ integral is
exactly the integral (\ref{eq:3.13}) with $u\rightarrow u'-y'$.  We can thus
write \begin{equation}
\tilde{V}_{\mu\nu'\rho'}^{(1)}=-2\pi^2\int d^4u'\,\left(\frac{(u'-y')_\sigma\,
(u'-y')_{\rho'}}{(u'-y')^6}-\frac{1}{2}\frac{\delta_{\sigma\rho'}}{(u'-y')^4}
\right)\frac{(u'-x')_\lambda}{(u'-x')^4}\,{\rm Tr}\,\gamma_5\gamma_\sigma
\gamma_\lambda\gamma_{\nu'}\gamma_\mu \label{eq:3.27} \, .\end{equation}

The $u'$ integral diverges logarithmically as $u'\rightarrow y'$ reflecting the
logarithmic divergence of individual vertex diagrams.  However we must now
add the contributions of the diagrams of Fig.~1c,d in which the vertex parts of $V^{(2)}$
and $V^{(3)}$ of (\ref{eq:3.11}-\ref{eq:3.12}) will appear (the latter with
modified photon propagator (\ref{eq:3.24})).  After the inversion process, one
finds that the $/\hspace{-2mm}x/x^4$ and $/\hspace{-2mm}v/v^4$ propagators
disappear, so that the $v'$-integrals are again those of (\ref{eq:3.14}) and
(\ref{eq:3.15}).  We now consider the sum of $V_\rho^{(1)}, V_\rho^{(2)}$ and
gauge-modified $V_\rho^{(3)}$ insertions in the two-loop diagrams.  It follows
by inspection of (\ref{eq:3.13}-\ref{eq:3.14}) that the net effect of the sum
is to reduce the coefficient of the $\delta_{\sigma\rho'}$ term in 
(\ref{eq:3.27})
by a factor of 2 giving a traceless tensor in those indices, so that the 
remaining convolution integral in $u'$ is convergent and may be read from
(\ref{eq:A10}).  The result for the net sum of vertex insertions at point $y$ of the
two-loop triangle is \begin{equation}
\tilde{V}_{\mu\nu'\rho'}=\frac{\pi^4}{4}\frac{{\rm Tr}\,\gamma_5\gamma_\mu
\gamma_{\rho'}\gamma_{\nu'}\,(/\hspace{-2mm}x'-/\hspace{-2mm}y')}{(x'-y')^4}
\, .\label{eq:3.28}\end{equation}

When combined with the prefactors in (\ref{eq:3.26}) and expressed as a
multiple of the one-loop amplitude one obtains \begin{equation}
V_{\mu\nu\rho}(0,x,y)=\frac{f^2}{64\pi^2}\,B_{\mu\nu\rho}(0,x,y)\, .
\label{eq:3.29} \end{equation}

The vertex insertions at points $x$ and $z$ must give the same contribution.
Again, we have studied the insertions at point $z$ using a conformal inversion
at $z\rightarrow0$.  A considerably more difficult set of integrals results
in the inverted variables, but the final result agrees with (\ref{eq:3.29}).

The various contributions to the order $g^3f^2$ amplitude must now be combined
with careful attention to combinatorics.  There is a factor of 3 from the
triangular symmetry, and a factor of two for opposite directions of fermion
charge flow for self-energy and vertex insertions, but not for the nonplanar
diagrams.  (From Fig.~1h we can see that the exchange $x\leftrightarrow
y$ produces a topologically equivalent diagram.)  Therefore our results
(\ref{eq:3.7}),(\ref{eq:3.23}) (with $(f^2-\frac{1}{2}g^2)\rightarrow f^2$) and 
(\ref{eq:3.29}) must be
added with weights \begin{equation}
3N_{\mu\nu\rho}+6(V_{\mu\nu\rho}+\Sigma_{\mu\nu\rho})=-\frac{3f^2}{8\pi^2}\,
\left(1-\frac{1}{4}-\frac{3}{4}\right)\,B_{\mu\nu\rho}\, ,
\label{eq:3.30} \end{equation}
showing that the net order $g^3f^2$ contribution to the gauge current
correlation function $\langle J_\mu(z)J_\nu(x)J_\rho(y)\rangle$ vanishes.

It will be useful for our treatment of more general chiral gauge theories to
give an alternative discussion of the integrals in the vertex insertion graphs.
We have seen that after conformal inversion the $\int d^4v'$ of the three
different vertex subgraphs can be read directly from (\ref{eq:3.13}-\ref
{eq:3.15}).  These expressions show that the dependence on the remaining
variable $u$ (which is transformed to $u'-y'$ in the two-loop graphs) is a 
superposition of an ``$s$-wave'' $\delta_{\sigma\mu}/u^4$ and a ``$d$-wave''
$(u_\sigma u_\mu-\frac{1}{4}\,\delta_{\sigma\mu}\,u^2)/u^6$ tensor form.  For
the pure gauge propagator (\ref{eq:3.24}), only the
vertex diagram $V_\mu^{(1)}$ has a $d$-wave, and the $s$-waves of $\int d^4v\,
V_\mu^{(1)}, \int d^4v\,V_\mu^{(2)}, \int d^4v\,V_\mu^{(3)}$ are in the ratio
of $1:-2:\frac{1}{2}\gamma$.  The final integral $\int d^4u'$ in the sum of
the three graphs diverges unless the net $s$-wave amplitude cancels, and this
selects the value $\gamma=2$ as the gauge parameter which makes the vertex
insertion subgraphs finite.

\setcounter{section}{3}
\mysection{The general $U(1)$ model.}

The action of this model has already been given in (\ref{eq:2.2}), and
the Feynman rules differ from the special case treated in Sec. 3 only in the
gauge vertex factors which now carry the chiral factors $-i\gamma_\mu(\alpha
L+\beta R)$.  We find it convenient to use the two coupling parameters $\beta$
and $\alpha$ and impose the relation $\beta-\alpha=1$ required for gauge
invariance selectively as necessary.

The major technical problem of the more general model is that there is no true
finite gauge due to the chiral gauge couplings.  This is immediately clear
from the one-loop self-energy amplitude \begin{equation}
\hat{\Sigma}(v-u)=\frac{1}{8\pi^2}\left[f^2+2g^2(\alpha^2R+\beta^2L)(1-\Gamma)
\right]\frac{{\textstyle /\hspace{-2mm}v-/\hspace{-2mm}u}}{{\textstyle(v-u)^6}}
+(aL+bR)\,/\hspace{-2mm}\partial_v\,\delta^4(v-u) \label{eq:4.1}\end{equation}
where, as in (\ref{eq:3.9}), we have included a possible local term.  One sees 
that for
$f\neq 0$ and $\alpha^2\neq\beta^2$ there is no value of the gauge parameter
$\Gamma$ which eliminates the ultraviolet singular $1/(v-u)^5$ factor.
Nevertheless we will see that there is an effective finite gauge which makes
the abnormal parity part of self-energy and vertex insertions on each line or 
each corner of the two-loop triangle finite.

Let us consider first all order $g^3f^2$ contributions to the current
correlator including all virtual Higgs graphs plus virtual photon graphs with
a pure gauge propagator similar to (\ref{eq:3.24}) but with $\gamma$ a 
numerical factor to be determined.  It is easy to obtain the
amplitude of the non-planar graph.  Clearing the chiral factors and comparing
with the previous case (\ref{eq:3.3}), one sees that we now have \begin{equation}
\hat{N}_{\mu\nu\rho}(0,x,y)=-4\alpha\beta\, N_{\mu\nu\rho} \, ,\label{eq:4.2}
\end{equation}
where $\,\hat{ }\,$ denotes the amplitude in the more general model.

For a given direction of fermion charge flow each vertex or self-energy
insertion graph contains both normal and abnormal parity amplitudes.  It
follows from Furry's theorem that the normal parity part cancels and the
abnormal parity part doubles in the sum of the two graphs with opposite
charge flow, so we can restrict our attention to the abnormal parity parts.
We let $\hat{V}_{\mu\nu\rho}^{(i)}\,(z,x,y)$ for $i=1,2,3$ denote the
abnormal parity part of the two-loop vertex graphs with vertex subgraph
$V_\rho^{(i)}(y,v,u)$ inserted at one corner.  The subgraph amplitudes are
given in (\ref{eq:3.10}--\ref{eq:3.12}) for $\beta=-\alpha=1/2$.

One can again manipulate chiral factors and compare with the previous case to
find, \begin{eqnarray}
\hat{V}_{\mu\nu\rho}^{(1)}\,(z,x,y)&=&-4\alpha\beta\,V_{\mu\nu\rho}^{(1)}
\,(z,x,y) \, .\nonumber \\
\hat{V}_{\mu\nu\rho}^{(2)}\,(z,x,y)&=&2(\alpha^2+\beta^2)\,V_{\mu\nu\rho}^
{(2)}\,(z,x,y)\, . \label{eq:4.3}\\
\hat{V}_{\mu\nu\rho}^{(3)}\,(z,x,y)&=&16(\beta^5-\alpha^5)\,V_{\mu\nu\rho}^{(3)}\,(z,x,y)\, .
\nonumber \end{eqnarray}
The relation $\beta-\alpha=1$ has been used in the first equality.

We now recall our discussion at the end of Section 3 of the integrals which
occur in the vertex insertion graphs after the conformal inversion is
implemented.  The integral $\int d^4v'$ gave the sum of $d$-wave and $s$-wave
tensors in $u'-y'$ for $V^{(1)}$ and pure $s$-waves for $V^{(2)}$ and $V^{(3)}$
with $s$-waves occurring in the ratio $1:-2:\frac{1}{2}\gamma$.  The gauge
parameter $\gamma$ must be chosen so that the net sum of the $s$-waves 
vanishes.
To implement this condition we must now weight the coefficients in (\ref
{eq:4.3}) by 1, -2 and $\frac{1}{2}\gamma$ thus obtaining $$
-4(\alpha^2+\alpha\beta+\beta^2)+8\gamma(\beta^5-\alpha^5)=0 $$
\begin{equation}
\gamma=\frac{(\beta^3-\alpha^3)}{2(\beta^5-\alpha^5)}
\label{eq:4.4}\end{equation}
for the choice of gauge fixing parameter which makes the sum of order 
$g^3f^2$ vertex insertion subgraphs at
each corner of the triangle finite.  For this choice the residual finite
contribution to $\langle J_\mu(z)J_\nu(x)J_\rho(y)\rangle$ comes just from the
$d$-wave tensor of $V^{(1)}$, and can be directly read from (\ref{eq:4.3}) as
\begin{eqnarray}
\hat{V}_{\mu\nu\rho}\,(z,x,y)&=&\hat{V}_{\mu\nu\rho}^{(1)}+\hat{V}_{\mu\nu\rho}
^{(2)}+\hat{V}_{\mu\nu\rho}^{(3)} \nonumber \\
 &=& -4\alpha\beta\,V_{\mu\nu\rho} \label{eq:4.5}\end{eqnarray}
where $V_{\mu\nu\rho}$ is given in (\ref{eq:3.29}).

Finally we must consider the self-energy insertion graphs.  We must check
that they are finite in the same gauge as the vertex insertion diagrams,
and we must determine the contribution of possible local terms in 
$\Sigma(u-v)$, {\em i.e.} $a$ or $b$ in (\ref{eq:4.1}).  To check finiteness
we consider the insertion of $\Sigma(v-u)$ of (\ref{eq:4.1}) into the two-loop
graph (Fig.~1e,f,g).  We move all chiral factors in the graph to the clockwise side
of the inserted $\Sigma(v-u)$.  This gives \begin{eqnarray}
(\alpha^3 R+\beta^3 L)\Sigma(v-u)&=&\frac{1}{8\pi^2}\left[f^2(\alpha^3 R+
\beta^3 L)-2g^2\Gamma(\alpha^5 R+\beta^5 L)\right]\frac{/\hspace{-2mm}v-
/\hspace{-2mm}u}{(v-u)^6} \, .\nonumber \\
 & &+(\alpha^3b\,R+\beta^3a\,L)/\hspace{-2mm}\partial_v\,\delta^4(v-u)
\label{eq:4.6} \end{eqnarray}
The sum over graphs with opposite direction of fermion charge flow selects
the abnormal parity part, namely \begin{eqnarray}
\frac{1}{16\pi^2}\left[f^2(\alpha^3-\beta^3)-2g^2\Gamma(\alpha^5-\beta^5)
\right]\,\gamma_5\,\frac{/\hspace{-2mm}v-/\hspace{-2mm}u}{(v-u)^6} 
\nonumber \\
+\frac{1}{2}(\alpha^3b-\beta^3a)\,\gamma_5\,/\hspace{-2mm}\partial_v\,
\delta^4(v-u)\, .
\label{eq:4.7}\end{eqnarray}
This effective self-energy must be multiplied by propagators for adjacent fermions and 
integrated $\int d^4v\,d^4u\,$.  The integral diverges unless the gauge
parameter is chosen so that the singular $1/(v-u)^5$ term in (\ref{eq:4.7})
cancels.  It is a relief, but hardly a surprise, to see that cancellation
occurs for the value of $\gamma$ given in (\ref{eq:4.4}), which also makes
vertex contributions finite.

We now see that, in the effective finite gauge, the abnormal parity part of
the self-energy insertion (Fig.~1e,f,g) involves only the local part of
$\Sigma(v-u)$ (Fig.~1g) given by the second term in (\ref{eq:4.7}).  The singularities
from the diagrams shown in Fig.~1e and Fig.~1f cancel.  The local term will now
be obtained from the Ward identity \begin{equation}
\frac{\partial}{\partial y_\rho}\,\hat{V}_\rho\,(y,v,u)=ig(\alpha R+\beta L)
[\delta^4(y-v)-\delta^4(y-u)]\hat{\Sigma}(v-u) \label{eq:4.8}\end{equation}
obtained by direct differentiation of the vertex graphs $\hat{V}_\rho^{(i)}$
of the model with general couplings.  The integrated form, which is the
generalization of (\ref{eq:3.19}), is
\begin{equation}
\frac{\partial}{\partial y_\rho}\int d^4v\,\hat{V}_\rho\,(y,v,0)=ig(\alpha R+
\beta L)\hat{\Sigma}(y) \, .\label{eq:4.9} \end{equation}

We now note that in the environment of the larger two-loop graphs 
(Fig.~1b,c,d),
all vertices at the $y$-corner of the triangle acquire the factor$\alpha^2 R
+\beta^2 L$ obtained by moving the $\alpha L+\beta R$ projectors at the $z$
and $x$ corners to the clockwise side of the point $v$.  We are thus 
specifically
interested in the abnormal parity part of the effective Ward identity
\begin{equation}
(\alpha^2R+\beta^2L)\frac{\partial}{\partial y_\rho}\int d^4v\,\hat{V}_\rho\,
(y,v,0)=ig(\alpha^3R+\beta^3L)\hat{\Sigma}(y) \, ,\label{eq:4.10} \end{equation}
and we observe that the chiral factor on the right side is exactly that of the
effective self-energy insertion in (\ref{eq:4.6}).

In the effective finite gauge  (\ref{eq:4.4}) the integral in (\ref{eq:4.10})
involves only the $d$-wave tensor from the $\hat{V}_\rho^{(1)}$ amplitude.
This contains an additional $\alpha L+\beta R$ chiral vertex factor, so the
coefficient of the abnormal parity part of (\ref{eq:4.10}) comes from 
$\alpha^2\beta\,R+\beta^2\alpha\,L$ and gives $-\frac{1}{2}\alpha\beta\,\gamma
_5$.  The value of the integral is then a factor of two times (\ref{eq:3.20})
(with $f^2-\frac{1}{2}g^2$ replaced by $f^2$ in (\ref{eq:3.20}) since we are
now considering the order $f^2$ terms only).  After computing the 
$\partial/\partial y_\rho$ divergence, as in
(\ref{eq:3.21}--\ref{eq:3.21.5}), the abnormal parity part of 
(\ref{eq:4.10}) reads,
after dropping the factor $ig$ on both sides, \begin{equation}
-\alpha\beta\frac{3f^2}{64\pi^2}\,\gamma_5\,/\hspace{-2mm}\partial\,\delta^4(y)
=(\alpha^3R+\beta^3L)\hat{\Sigma}(y)\,|_{\stackrel{{\scriptstyle {\rm abn.}}}
{{\scriptstyle {\rm par.}}}} \,\, .\label{eq:4.11}\end{equation}
This equation gives $\gamma_5$ times the effective self-energy including
chiral factors from the corners of the two-loop triangle.  With a little
thought we can then see that each self energy graph of the general chiral theory
is related to (\ref{eq:3.23}) by \begin{equation}
\hat{\Sigma}_{\mu\nu\rho}\,(z,x,y)=-4\alpha\beta\,\Sigma_{\mu\nu\rho}\,(z,x,y)
\, .\label{eq:4.12}\end{equation}
(An extra negative sign has been gained by moving the $\gamma_5$ past
the propagator $S(z-v)$ to its original position in the trace of the two-loop
graphs.)

The results (\ref{eq:4.2}),(\ref{eq:4.5}) and (\ref{eq:4.12}) show that all
contributions to the abnormal parity part of the $\langle J_\mu J_\nu J_\rho
\rangle$ correlators in the $U(1)$ Higgs theory with general chiral couplings
are a uniform factor of $-4\alpha\beta$ times the corresponding contributions
in Section 3.  Thus the sum of all order $g^3f^2$ terms in the correlation
function vanishes.

We now discuss the order $g^5$ virtual photon contributions.  We see from
(\ref{eq:4.1}) and (\ref{eq:3.16}) that the Landau  gauge is a true finite
gauge which makes the one-loop vertex and self-energy graphs entirely
finite.  This makes the argument simpler.  We first note that all order
$g^5$ graphs contain the chiral factor $(\alpha L+\beta R)^5$, whose abnormal
parity part is just $\frac{1}{2}(\beta^5-\alpha^5)$ times the corresponding
graph in the QED correlator studied in \cite{BJ}.  Further, when chiral
factors are extracted from the Ward identity (\ref{eq:4.9}) one can see that it
coincides with the QED Ward identity used by \cite{BJ} to determine the
local part of the self-energy.  So the analysis of \cite{BJ} applies in its
entirety and shows that the two-loop $\langle J_\mu J_\nu J_\rho\rangle$
correlator also vanishes in the chiral $U(1)$ model.

\setcounter{section}{4}
\mysection{Standard model anomalies.}

We next calculate the two loop anomalies in the $SU(3)\times SU(2)\times U(1)$
gauge theory with one generation of quarks and leptons.  All fields are
massless.

The Euclidean Lagrangian is \begin{eqnarray}
{\cal L}&=&\frac{1}{4}\stackrel{\rightarrow}{F^{\mu\nu}}\cdot\stackrel{\rightarrow}{F^{\mu\nu}}+\sum_
{\stackrel{{\scriptstyle{\rm quarks}}}{{\scriptstyle {\rm leptons}}}}
\overline{\psi}_{iI}\gamma^\mu\Big{[}\partial_\mu-i g_3\,\delta_{ij}\delta_{\psi,{\rm quark}}\,
T^A_{IJ}\, G^A_\mu \nonumber \\
 & &-\frac{i g_2}{2}\delta_{IJ}\,\tau^a_{ij}W^a_\mu L-\frac{ig_1}{2}
\delta_{ij}\delta_{IJ}\left(Y_LL+Y_RR\right)B_\mu\Big{]}\psi_{jJ} \nonumber \\
 & &+\phi^\dagger\left(\stackrel{\leftarrow}{\partial}_\mu+\frac{ig_1}{2}\stackrel{\rightarrow}{\tau}\cdot \stackrel{\rightarrow}{W}_\mu
+\frac{ig_1}{2}B_\mu\right)\left(\partial_\mu-\frac{ig_2}{2}\stackrel{\rightarrow}{\tau}\cdot \stackrel{\rightarrow}{W}_\mu
-\frac{ig_1}{2}B_\mu\right)\phi \nonumber \\
 & &-f_l\left(\overline{l}_i\,\phi_iR\,e+\overline{e}\,\phi_i^\dagger L\,l_i\right)
-f_d\left(\overline{q}_i\,\phi_iR\,d + \overline{d}\,\phi_i^\dagger L\,q_i\right) \nonumber \\
 & &-f_u\left(\overline{q}_i\left(i\tau^2\right)_{ij}\phi_j^\dagger \,R\,u + \overline{u}\, \phi_j
\left(i\tau^2\right)_{ji}L\,q_i\right)-\frac{1}{4}\lambda(\phi^\dagger\phi)^2
\, .\label{eq:smlagrangian} \end{eqnarray}
$\tau^a_{ij}$ are the Pauli matrices and $T^A_{IJ}$ the Gell-Mann matrices.
$B_\mu$ is the abelian gauge boson.  The non-abelian gauge bosons $W^a_\mu$
and $G^A_\mu$ will be referred to collectively as gluons.  Lower case Latin
indices ($i,j,a$) refer to $SU(2)$; upper case Latin indices ($I,J,A$) refer
to $SU(3)$.  $l_i$ is the lepton $SU(2)$ doublet ($\nu,e$); $q_i$ is the quark
doublet ($u,d$); $\phi_i$ is the Higgs doublet ($\phi^+,\phi^0$).  The leptons
and Higgs are singlets under $SU(3)$, while the quarks are triplets.  The
$U(1)$ hypercharges of the standard model matter fields are tabulated below.
The one-loop anomalies are easily shown to be absent with this assignment.

\begin{table}[h]\begin{center}
\begin{tabular}{c|c|ccr}
\multicolumn{2}{r|}{$Y_L$} & $Y_R$&\multicolumn{2}{c}{ }  \\ \cline{2-3}
$\nu$ & -1 & 0 & \\ \cline{1-3}
$e$ & -1 & -2& &\hspace{0.2in}$\phi_i:$\ Y=1 \\ 
\cline{1-3}
$u$ & 1/3 & 4/3 & \\ \cline{1-3}
$d$ & 1/3 & -2/3 & \\ \end{tabular}
\end{center}
\caption{Fermion and Higgs hypercharges.}
\end{table}

The diagrams which contribute to the two-loop anomalies are essentially the
same as those considered in Section 4, except that the fermion and gauge lines
now carry group indices.  We structure the calculation by comparison to the
previous abelian case with pure axial gauge coupling.  There are in addition 
several
diagrams (Fig.~2b,c,d) not considered previously involving vertices with three
non-Abelian gauge bosons, but we show that they do not contribute because the
anomaly vanishes at the one-loop level.  The extension to $N$ generations of
quarks and leptons with unitary CKM mixing matrix $M_{ij}$ for the quarks adds
no new complications.  Each gauge boson exchange diagram gets the factor
${\rm Tr}\,M^\dagger M=N$, so each two-loop contribution to the correlator is 
identical to that given below times $N$.

In the standard model there is a potential anomaly for each choice of the
three currents in the correlator.  However, each diagram is proportional to
the trace of the product of the three corresponding group generators.  Since
the trace of any (non-$U(1)$) generator is zero, the only three-current correlators which
are potentially nonvanishing are $\langle J_1J_1J_1\rangle , \langle J_2J_2J_2
\rangle, \langle J_2J_2J_1\rangle ,$ and $\langle J_3J_3J_1\rangle .$
The $SU(3)$ current is non-chiral; hence the $\langle J_3J_3J_3\rangle$ correlator
is not anomalous.  Furthermore, after summing over the two
directions of fermion flow, each diagram is proportional to the group theory
$d$-symbols, ${\rm Tr}\, T^a\{T^b,T^c\}.$  Since the $d$-symbols vanish for $SU(2)$
the correlator $\langle J_2J_2J_2\rangle$ vanishes, as well.

The contributions to the abnormal parity part of the two-loop correlators from 
$SU(2)$ and $SU(3)$ gluon exchange vanish.  All such contributions are 
proportional to the group theoretic factor which vanishes by the same condition
which enforces the cancellation of the corresponding one-loop anomaly.  We 
take the gauge parameter $\Gamma$ for the $U(1)$ vector
boson as the Landau gauge value 1 plus a term proportional to the Yukawa
couplings which we will discuss below.  One can then show that the Landau
gauge contributions to all vertex and self-energy insertion diagrams for the
three-point current correlators vanish by the results of \cite{BJ}.

The treatment of the ambiguous local part of the self-energy via the Ward
identities will be treated in a different (but equivalent) way from Section 4.
We always check that in each gauge invariant sector of the calculation there
is a choice of the $U(1)$ vector boson gauge parameter $\Gamma$ which makes the 
would be divergent $s$-wave vertex contributions and $1/(u-v)^5$ self-energy
contributions to the abnormal parity parts of the correlators simultaneously
finite (and zero).  It is then justified to keep only the $d$-wave parts of
the once integrated $\int d^4v'$ vertex subgraphs, and the same for the
integrated Ward identity used to determine the local part of the self-energy.

\subsection{$\langle J_1 J_1 J_1\rangle$}
The $\langle J_1J_1J_1\rangle$ calculation in the standard model is
directly analogous to that of the $\alpha L+\beta R$ theory considered in
Section 4.  The only
difference is the sum over the $SU(2)$ and $SU(3)$ indices.  We first check
that the effective finite gauge mechanism works as in the previous case.  It
is sufficient that such a gauge exist for each gauge invariant sector of the
calculation.  Since the leptons and quarks do not mix to this order we find
separate values for the gauge parameter $\Gamma$ which make the lepton and
quark vertex and self-energy insertions finite.  
We calculate the lepton contribution below.  The quark calculation is
analogous, with the charges replaced appropriately.  

Including the relevant
chirality factors and summing the contributions from internal electron and 
neutrino propagation, we easily find
that the abnormal parity parts of $\hat{V}^{(1)}$, $\hat{V}^
{(2)}$ and $\hat{V}^{(3)}$ (Fig.~1b,c,d) are, with the point $z$ taken 
to zero,
\begin{eqnarray}
\hat{V}_{\mu\nu\rho}^{(1)}(0,x,y)=\left(\frac{-1}{2}\right)^3\cdot 2\cdot (-4)
\,Y_L^{(e)}Y_R^{(e)}\,V_{\mu\nu\rho}^{(1)}(0,x,y)\, ,
\label{eq:1-higgs1} \\
\hat{V}_{\mu\nu\rho}^{(2)}(0,x,y)=\left(\frac{-1}{2}\right)^3\cdot 2\cdot 2
\,(Y_L^{(e)\, 2}+Y_R^{(e)\, 2})\,V_{\mu\nu\rho}^{(2)}(0,x,y)\, ,
\label{eq:2-higgs1} \\
\hat{V}_{\mu\nu\rho}^{(3)}(0,x,y)=\left(\frac{-1}{2}\right)^5\, (-16)\sum_{
\stackrel{{\scriptstyle {\rm neutrino}}}
{{\scriptstyle {\rm electron}}}}(Y_R^5-Y_L^5)\,V^{(3)}_{\mu\nu\rho}\,  .
\label{eq:gauge1} \end{eqnarray}
where $V^{(1)}, V^{(2)}$ and $V^{(3)}$ are the same diagrams in the pure axial
$U(1)$ theory considered in Section 3.  The left-handed electron and neutrino 
form an $SU(2)$ doublet, so they have the same hypercharge $Y_L^{(e)}$.  The 
superscript $(e)$ on the hypercharges denotes explicitly that only the 
right-handed electron hypercharge contributes. Of course, this 
comment is trivial since there 
is no right-handed neutrino in the theory. but equations (\ref{eq:1-higgs1}--
\ref{eq:gauge1}) are valid for the quark contributions as well with the
appropriate replacement of lepton labels by quark labels.  For example, the
contribution from the $f_d$ Yukawa coupling is obtained from (\ref{eq:1-higgs1}
--\ref{eq:2-higgs1}) by the replacement $Y_{L,R}^{(e)}\rightarrow Y_{L,R}^{
(d)}$.
The factors of $(-1/2)^3$ and $(-1/2)^5$ come
 from the 
difference in the definition of charge from the previous case, and one
factor of $2$ in (\ref{eq:1-higgs1}) and (\ref{eq:2-higgs1}) is due to the 
$SU(2)$ trace over the electron and neutrino.
The condition for cancellation of divergent $s$-wave integrals in the sum of
$V_{\mu\nu\rho}^{(1)}, V_{\mu\nu\rho}^{(2)}$ and $V_{\mu\nu\rho}^{(3)}$ 
can now be written as the sum of the hypercharge coupling factors in 
(\ref{eq:1-higgs1}-\ref{eq:gauge1}), each weighted by the factors 
$-1:2:-\gamma/2$ of
the $s$-wave integrals in (\ref{eq:3.13}-\ref{eq:3.15}).  This gives 
the effective finite gauge condition
\begin{equation}
Y_L^{(e)}Y_R^{(e)}+(Y_R^{(e)\, 2}+Y_L^{(e)\,2})+\frac{\gamma}{4}\sum (Y_R^5-Y_L^5)=0\, .
\end{equation}
We checked that this condition agrees with that obtained from insisting that
the sum of the self energy diagrams is finite and indeed zero up to the
ambiguity of local terms.  Note that even if we were to choose the gauge
parameter for the nonabelian gauge fields to be other than the Landau gauge
value $\Gamma=1$, the additional contributions to vertex and self-energy
insertions after summing over quarks and leptons are proportional to the 
one-loop anomaly, which is zero.

It is easy to check that the nonplanar diagram is multiplied by the same factor
as $\hat{V}^{(1)}$ when compared to the pure axial $U(1)$ case, (\ref{eq:3.1}):
\begin{equation}
\hat{N}_{\mu\nu\rho}(0,x,y)=Y_L^{(e)}Y_R^{(e)}\,N_{\mu\nu\rho}(0,x,y)\, .
\end{equation}

Next we calculate the local part of the self energy which contributes when
inserted in the fermion triangle (Fig.~1g) in the effective finite gauge.
If we denote by $V_i^\mu(z,u,v)$ the lepton vertex ($i$=electron or neutrino) with
charge flowing from $v$ to $u$ and by $\Sigma_1(u-v)$ the self energy with charge
flowing in the same direction,
then the relevant Ward identity is, as in (\ref{eq:4.8}),
\begin{equation}
\partial_\mu^zV_i^\mu(z,u,v)=\frac{ig_1}{2}\left[\delta^4\left(z-v\right)-
\delta^4(z-u)\right]\left(Y_R^iL+Y_L^iR\right)\Sigma_1^i(u-v)\, ,
\end{equation}
This can be integrated to give \begin{equation}
\int d^4u\,\partial_\mu^z\,V_i^\mu=-\frac{ig_1}{2}(Y_R^iL+Y_L^iR)\,\Sigma_1(z-v)\, . \label{eq:WI1}
\end{equation}

We calculate the $d$-wave (traceless) part of the integrated vertex as before,
since the $s$-wave contributions and the corresponding $1/z^5$ part of the
self-energy vanish in the effective finite gauge.  Again, the 
only $d$-wave contribution is from Fig.~1b with
a single Higgs exchanged.  We find that \begin{equation}
\int d^4u\,\partial^z_\mu\, V_i^\mu(z,u,v)\,\stackrel{d-{\rm wave}}{=}\,\frac
{3ig_1f_l^2}{16(4\pi^2)}(2Y_L^{(e)}L\delta_{i(e)}+Y_R^{(e)}R)/\hspace{-2mm}
\partial_z\delta^4(z-v)\, . \label{eq:dwave1}\end{equation}
Projecting out the left and right handed pieces of (\ref{eq:dwave1}) and
comparing with the Ward identity (\ref{eq:WI1}), we find for the relevant
contribution of the local part
of the self energy \begin{equation}
\Sigma_1^i(z-v)=\frac{-3f_l^2}{16(4\pi^2)}\left(2\frac{Y_L^{(e)}}{Y_R^{(e)}}L
\delta_{i(e)}+\frac{Y_R^{(e)}}{Y_L^{(e)}}R\right)/\hspace{-2mm}\partial
\delta^4(z-v)\, . \label{eq:5.10}\end{equation}
The $\delta_{i(e)}$ means that the right handed electron (but not neutrino) 
flows through $\Sigma$.

We insert this self energy into the fermion triangle (Fig.~1g).  After 
pulling together chiral factors and summing over the electron and neutrino
we indeed find a multiplicative factor 
$Y_L^{(e)}Y_R^{(e)}$ times the result in the pure axial
case considered in Section~3.  \begin{equation}
\hat{\Sigma}_{\mu\nu\rho}(0,x,y)=Y_L^{(e)}Y_R^{(e)}\,\Sigma_{\mu\nu\rho}(0,x,y)\, ,
\end{equation}
with $(f^2-\frac{1}{2}g^2)\rightarrow f^2$ in $\Sigma_{\mu\nu\rho}$ 
(\ref{eq:3.23}).  

Thus the vertex, non-planar and self-energy diagrams are each multiplied by an 
overall factor of $Y_L^{(e)}
Y_R^{(e)}$ times the corresponding diagrams in the pure axial $U(1)$ case
treated in Section 3.  Note that with $\alpha,\beta\rightarrow Y_L^{(e)},
Y_R^{(e)}$, this result is identical to that of Section~4 up to a
factor of (-4) which comes from the difference in definition of $U(1)$ charge
and the sume over electron and neutrino.
So the sum of all virtual lepton
contributions to the correlator $\langle J_1^\mu J_1^\rho J_1^\nu\rangle$
vanishes as in the previous case.

The quark
contributions vanish similarly.  Note that while the quarks have two different 
Yukawa couplings, to this order these couplings do not mix because
of the relative chirality flip between the two couplings.  Furthermore,
 while the up type Yukawa coupling
$f_u$ in the Lagrangian (\ref{eq:smlagrangian}) seems more complicated than
the down or electron type,  we can redefine the Higgs
field to be the $SU(2)$ conjugate $\phi'_i=(i\tau_2)_{ij}\,\overline{\phi}_j$, 
and then the
kinetic term in terms of the conjugated Higgs field looks identical to the
previous case except for a change of sign of the Higgs charge.  Then the 
calculation is identical to the previous case,
as long as we remember that for the up quark $Y_R-Y_L=1,$ while for the leptons
and down quark $Y_L-Y_R=1$ as required for $U(1)$ gauge invariance of the
Yukawa couplings.  

Hence the total contribution from the gauge fields 
and each of the three Yukawa couplings to the two-loop correlator vanishes, as
expected.

\subsection{$\langle J_2J_2J_2\rangle$}
We noted earlier that the correlator $\langle J_2J_2J_2\rangle$ vanishes by
virtue of the fact that the $SU(2)$ $d$-symbols are all zero.  However, we
expect that the vanishing of the anomaly should not depend on the gauge
group, so long as the quarks and leptons are in a representation for which
the one-loop anomaly vanishes.  We therefore check that even if we neglect the
fact that the group theoretic $d$-symbols are zero the correlator still 
vanishes.  We only require that the trace of the $d$-symbols over the left-%
handed fermion representations in the theory vanishes.

The calculation is remarkably simple as a result of the left-handedness of
the $SU(2)$ current.  Since the Yukawa couplings in the Lagrangian
(\ref{eq:smlagrangian}) 
(\ref{eq:smlagrangian}) always connect a left-handed
field to a right-handed one, it is easy to see that both the nonplanar
diagram and the vertex insertion with a single Higgs exchanged vanish.
The contribution from
the self energy is determined via the $SU(2)$ Ward identity and vanishes
because the $d$-wave part of the vertex contribution vanishes.  
Since these are the
only contributions to the correlator in the effective finite gauge, which
we also checked to exist, the correlator vanishes to two-loops.

It appears at first that there are additional diagrams (Fig.~2b,c,d) which we
have not calculated, but they are all proportional to the one-loop anomaly. 
Fig.~2b contains the one-loop
triangle, so it is immediately proportional to the one-loop anomaly.  The
group theory of the other diagrams is easy to work through.  The three-gauge
vertex is proportional to the $SU(2)$ structure constant $f^{abc} (=
\varepsilon^{abc})$.  Summing
over the two directions of fermion flow, we see that the abnormal
parity  part of each of the diagrams (Fig.~2c,d) is proportional to ${\rm Tr}\,f^{ade}(\tau^b\tau^c\tau^e\tau^d-
\tau^d\tau^e\tau^c\tau^b)=
{\rm Tr}\,f^{ade}f^{edg}\tau^b\{\tau^c,\tau^g\} = -2{\rm Tr}\,
d^{abc}$.  So each two-loop diagram in question contains the factor ${\rm Tr}\,
d^{abc}$, where the trace sums over all left-handed fermions in the theory.
The condition ${\rm Tr}\, d^{abc}=0$ also makes the one-loop anomaly vanish,
as promised above.  One can easily check that the same effect 
occurs for the contribution
of the analogous diagrams to the other correlators discussed below.  In both
of the cases treated below in Sections~5.3 and 5.4, Fig.~2c,d are proportional 
to the factor which makes the corresponding one-loop anomaly vanish.

\subsection{$\langle J_1(z)J_2(x)J_2(y)\rangle$}
The calculation of the $\langle J_1J_2J_2\rangle$ and $\langle J_1J_3J_3\rangle$
correlators is complicated by the fact
that the local part of the self energy calculated from the $U(1)$ Ward identity
and the $d$-wave part of the $U(1)$ vertex is not consistent with that
calculated via the $SU(2)$ ($SU(3)$) Ward identity and the $d$-wave part of the
$SU(2)$ ($SU(3)$)
vertex.  This is not surprising since the vertex is also ambiguous up to
a local term as a result of renormalization.  We are free to fix the arbitrary
coefficient of the local self energy, as long as we concurrently add local 
parts to the 
vertices to make them consistent with the Ward identities.

We will use the local part of the self energy calculated in section Section~%
5.1 and modify the local $SU(2)$ vertex to make it consistent with this
choice.  The $U(1)$ vertex is unmodified.  Again, we calculate only the lepton
contribution here.  The quark contribution is analogous.

First we find the effective finite gauge for this calculation.
Note that the $d$-wave contribution to the integrated
$SU(2)$ vertex vanishes.  The left-handed fields do not contribute because
at the Yukawa vertex they become right-handed fields which do not couple to the
$SU(2)$ current.  The right-handed fields do not contribute because they are
$SU(2)$ singlets; alternatively, their contribution is proportional
to ${\rm Tr}\,\tau^a=0$.  The only nonvanishing two-loop diagram of the form
Fig.~1b
is from the $U(1)$ vertex, which is inserted at the point $z=0$ in the two-loop
triangle.  It contributes \begin{equation}
\hat{V}_{\mu\nu\rho}^{(1\,z)\,ab}=\frac{1}{2}Y_R^{(e)}{\rm Tr}\, 
\tau^a\tau^b\,V^{(1)}_{\mu\nu\rho}\, . 
\label{eq:1-higgs2}\end{equation}
The label $z$ on $\hat{V}^{(1\,z)}$ denotes that the vertex is placed at the
$U(1)$ corner, which is chosen to lie at point $z$.

In diagrams of the form Fig.~1c, the two-higgs vertex can lie at
the $U(1)$ corner or either of the $SU(2)$ corners of the triangle.  When it
lies at the $U(1)$ corner it contributes \begin{equation}
\hat{V}_{\mu\nu\rho}^{(2\,z)\,ab}=-\frac{1}{4}\, {\rm Tr} \,\tau^a\tau^b\,
V_{\mu\nu\rho}^{(2)}\, .
\label{eq:2-higgs21}\end{equation}
When it lies at one of the $SU(2)$ vertices it contributes \begin{equation}
\hat{V}_{\mu\nu\rho}^{(2\,x)\,ab}=\hat{V}_{\mu\nu\rho}^{(2\,y)\,ab}=
-\frac{1}{4}\,Y_L^{(e)}\,{\rm Tr}\,\tau^a\tau^b\, V_{\mu\nu\rho}^{(2)}\,
 .
\label{eq:2-higgs22}\end{equation}
The contribution to the vertex diagrams from the effective $U(1)$ gauge boson
 propagator
is the same at each corner of the triangle.  It is \begin{equation}
\hat{V}_{\mu\nu\rho}^{(3\,z)\,ab}=\hat{V}_{\mu\nu\rho}^{(3\,x)}=
\hat{V}_{\mu\nu\rho}^{(3\,y)\,ab}=-Y_L^{(e)\, 3}\,{\rm Tr}\,\tau^a\tau^b\,
V_{\mu\nu\rho}^{(3)}\, .  
\label{eq:photon2}
\end{equation}
Including the relative coefficients of $V^{(1)}, V^{(2)}$ and $V^{(3)}$ in 
(\ref{eq:1-higgs2}--\ref{eq:photon2}) in the ratio $-1:2:-\gamma/2$ as
before and summing over the three corners of the triangle,
we find the condition for the divergent ($s$-wave) parts of the vertex 
diagrams to cancel, \begin{equation}
1-\frac{\gamma}{2}\,Y_L^{(e)\,2}=0 \, . \end{equation}
Although we obtained this result by summing over the three corners of the
triangle, the same condition makes the $U(1)$ and $SU(2)$ vertex contributions
separately finite.  This is expected because they
are related to the same self-energy insertion graphs by Ward identities, as in
(\ref{eq:4.10}).

We next consider the contribution from the local $SU(2)$ vertex correction 
to the two-loop correlator.
The relevant $SU(2)$ Ward identity is \begin{equation}
\partial^z_\mu \,V_{ji}^{a\,\mu}(z,u,v)=\left(\frac{-ig_1}{2}\right)\tau^a_{ji}
\left[\delta^4(z-u)-\delta^4(z-v)\right]\Sigma(u-v)L\, , \label{eq:SU(2)WI}
\end{equation} 
where the $SU(2)$ charge flows from $i$ to $j$ and $\Sigma(u-v)$ is the left-handed
part of the self energy (which is the same for the electron and neutrino).

In the effective finite gauge the sum of the electron and neutrino 
contributions to the vertex and self-energy insertions is finite.  
Equivalently, there are no $s$-wave parts of vertex insertions, so we can
confine ourselves to just the $d$-wave part of the Ward identity and speak
separately about the electron and neutrino.
However, the $SU(2)$ vertex insertion has no Higgs exchange diagram (Fig.~1b)
and thus no $d$-wave part.  Thus, if we were to use the $SU(2)$ Ward identity to derive the consistent
local part to the self energy it would vanish as well.  Thus the self-energy
(\ref{eq:5.10}) is not consistent with (\ref{eq:SU(2)WI}).  Hence we modify the
vertex by a local contribution as discussed.  We add to the $SU(2)$ 
vertex a local part of the form \begin{equation}
\Delta V^{a\,\mu}_{ji}(z,u,v)=\frac{ig_2}{2}Z\tau^a_{ji}\,\gamma^\mu L\,\delta^4(z-u)\,
\delta^4(z-v)\, .  \end{equation}
Its divergence is easily calculated to be \begin{equation}
\frac{\partial}{\partial z^\mu}\,\Delta V^{a\,\mu}_{ji}(z,u,v)=\frac{ig_2}{2}Z\tau^a_{ji}
[\delta^4(z-v)-\delta^4(z-u)]R\, /\hspace{-2mm}\partial_v\,\delta^4(v-u) \, . 
\end{equation}
With the self energy given in eq. (\ref{eq:5.10}), the Ward identity
(\ref{eq:SU(2)WI}) determines the parameter $Z$ to be \begin{equation}
Z=\frac{-3f^2}{16(4\pi^2)}\frac{Y_R^{(e)}}{Y_L^{(e)}} \, . \end{equation}
Inserting this vertex renormalization at either of the $SU(2)$ corners of the
one-loop lepton triangle (Fig.~3a,b), we immediately find a result proportional
to the one-loop amplitude, whose abnormal parity part $\Delta V_{\mu\nu\rho}^{%
ab}$ is:

\begin{figure}\begin{center}
\PSbox{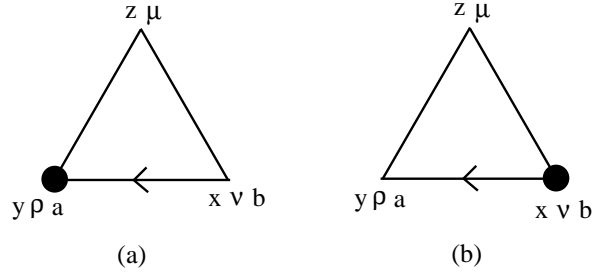}{3.1in}{1in}
\end{center}
\caption{Contributions to $\langle J_1^\mu(z)J_2^{a\,\rho}(y)J_2^{b\,\nu}
(x)\rangle$ from the $SU(2)$ vertex renormalization.  In Section~5.4 we instead
renormalize the $U(1)$ vertex.}
\end{figure}

\begin{eqnarray}
\Delta V_{\mu\nu\rho}^{ab}(0,x,y)&=&\frac{1}{2}\frac{(-3f_l^2)}{16(4\pi^2)}\,
Y_R^{(e)}\,{\rm Tr}\,\tau^a\tau^b
\,B_{\mu\nu\rho}(0,x,y) \nonumber \\ 
 &=&-\frac{1}{2}Y_R^{(e)}\, {\rm Tr}\, \tau^a\tau^b\, \Sigma_{\mu\nu\rho}
(0,x,y)\, ,  \label{eq:vertex2}\end{eqnarray}
where $B_{\mu\nu\rho}$ and $\Sigma_{\mu\nu\rho}$ are given in (\ref{eq:3.5})
and (\ref{eq:3.23}) (with $(f^2-\frac{1}{2}g^2)\rightarrow f^2$), respectively.
There is a similar contribution when the vertex
correction is placed at either of the $SU(2)$ corners of the lepton triangle.

The contribution from the diagram (Fig.~1g) with the self
energy (\ref{eq:5.10}) inserted at any of the three legs of the triangle is 
easily calculated to be
\begin{equation}
\hat{\Sigma}_{\mu\nu\rho}^{ab}=\frac{1}{2}Y_R^{(e)}\,{\rm Tr}\,\tau^a\tau^b\,\Sigma
_{\mu\nu\rho}\, . \label{eq:self-energy2}
\end{equation}

The nonplanar diagram with the Higgs current placed 
at the $U(1)$ vertex vanishes because the chiral projectors at the Yukawa
vertices annihilate the propagating fermions.  The
abnormal parity contribution from each of the two remaining nonplanar diagrams 
is 
\begin{equation}
\hat{N}_{\mu\nu\rho}^{(x)\,ab}=\hat{N}_{\mu\nu\rho}^{(y)\,ab}=
\frac{1}{4}Y_R^{(e)}\, {\rm Tr}\,\tau^a\tau^b\,N_{\mu\nu\rho}\, .
\label{eq:nonplanar2}\end{equation}

Recall again that in the pure axial case the relative contributions of the
self-energy, vertex and nonplanar graphs were in the ratio $3:1:-4$, summing
to zero.  Adding the contributions in the effective finite gauge from 
(\ref{eq:1-higgs2}) and (\ref{eq:vertex2}--\ref{eq:nonplanar2}) gives
the now familiar result: \begin{equation}
\Big[ \underbrace{\left(-\frac{{\scriptstyle 1}}{{\scriptstyle 2}}\cdot 2+\frac
{{\scriptstyle 1}}{{\scriptstyle 2}}\cdot 3\right)\cdot
3}_{\stackrel{\scriptstyle{\rm vertex\,correction}}{{\scriptstyle{\rm + self\,energy}}}}+
\underbrace{\frac{{\scriptstyle 1}}{{\scriptstyle 2}}\cdot 1}_{{\scriptstyle
{\rm vertex}}}+\underbrace{\frac{{\scriptstyle 1}}{{\scriptstyle 4}}\cdot 2
\cdot({\scriptstyle -}4)}_{{\scriptstyle{\rm nonplanar}}}\Big]\,\, Y_R^{(e)}\,
{\rm Tr}\,\tau^a\tau^b \,=0 . \end{equation}

\subsection{$\langle J_1(z)J_3(x)J_3(y)\rangle$}
We proceed as in Section 5.3.  The difference here is that $SU(3)$ couples only to
quarks and the coupling is nonchiral.  First we check that the effective 
finite gauge mechanism works in this case.  Again, the two quark Yukawa
couplings $f_d$ and $f_u$ contribute independently at the two-loop level.
(They do not mix because of the relative chirality flip between
 the two couplings, as can be seen by trying to draw a two-loop Higgs exchange
diagram with both Yukawa couplings).
Furthermore, the contribution from $f_u$ is obtained from that of $f_d$ by the
interchange of $Y_L$ and $Y_R$.  We calculate the contribution from the $f_d$
coupling here.

With the one-loop Higgs exchange vertex inserted at the $U(1)$ corner 
(Fig. 1b) we get 
\begin{equation}
\hat{V}_{\mu\nu\rho}^{(1\,z)\,AB}=2\cdot 2\,{\rm Tr}\,T^AT^B\,(Y_R^{(d)}-Y_L^{
(d)})\,V_{\mu\nu\rho}^{(1)}
=-4\,{\rm Tr}\,T^A
 T^B\,V_{\mu\nu\rho}^{(1)}\, . \label{eq:1-higgs31}
\end{equation}
At each of the $SU(2)$ vertices it contributes the same except for a negative
sign
which can be traced to the vector nature of the $SU(3)$ coupling,
\begin{equation}
\hat{V}_{\mu\nu\rho}^{(1\,x,y)\,AB}=-2\cdot 2\,{\rm Tr}\,T^AT^B\,(Y_R^{(d)}-Y_L^{
(d)})\, V_{\mu\nu\rho}^{(1)}
=4\,{\rm Tr}\,T^A
T^B\,V_{\mu\nu\rho}^{(1)}\, . \label{eq:1-higgs33}
\end{equation}

Since the Higgs is an $SU(3)$ singlet the only vertex diagram including the
Higgs current (Fig.~1c) is from the $U(1)$ corner of the triangle.
It contributes \begin{equation}
\hat{V}_{\mu\nu\rho}^{(2\,z)\,AB}=-2\cdot 2{\rm Tr}\,T^AT^B\, V_{\mu\nu\rho}^{(2)}
\, . \end{equation}

The $U(1)$ gauge boson exchange diagram (Fig.~1d) contributes at each corner
of the triangle \begin{equation}
\hat{V}_{\mu\nu\rho}^{(3\,z,x,y)\,AB}=-2\sum_{{\rm quarks}}(Y_L^3-Y_R^3)\,{\rm Tr}\,
T^AT^B\,V_{\mu\nu\rho}^{(3)}\, .
\end{equation}

Recalling that in the previous case the diagrams of Fig.~1b,c,d appeared in 
the ratio $1:-2:\gamma/2$, we find the effective
finite gauge condition either by summing over the three corners of the triangle
or by summing the contributions at any particular corner.  \begin{equation}
1+\frac{\gamma}{4}\sum_{{\rm quarks}}(Y_R^3-Y_L^3)=0\, .\end{equation}
One can easily check that the same condition follows from making the singular
$1/(v-u)^5$ parts of the self-energy cancel.

For aesthetic reasons we choose not to introduce a parity non-conserving 
correction to the $SU(3)$ fermion vertex but rather introduce a correction to 
the ambiguous part of the $U(1)$ vertex.  Then in the self energy insertion
Fig.~1g we determine the ambiguous local part of the self energy
via the relevant $SU(3)$ Ward identity and the $d$-wave part of the integrated
$SU(3)$ quark vertex.  The relevant identity is \begin{equation}
\partial_\mu^z\,V_{JI}^{A\,\mu}=ig_3[\delta^4(z-v)-\delta^4(u-z)]T^A_{JI}\,
\Sigma_3(u-v)\, ,
\end{equation}
where the notation is as before, with $I$ and $J$ $SU(3)$ indices.  We
calculate the $d$-wave part of the vertex $V_{JI}^{A\,\mu}$ to determine the
relevant ambiguous part of the self energy which contributes when inserted
in the fermion triangle.  We find \begin{equation}
\Sigma_3^i(z-v)=\frac{-3f_d^2}{16(4\pi^2)}(2L\delta_{i(d)}+R)/\hspace{-2mm}
\partial_z\,\delta^4(z-v) \, . \end{equation}
where the $\delta_{i(d)}$ means that the right handed down (but not up) quark
receives a self-energy contribution.
Inserting this self energy in the fermion triangle Fig.~1g gives
\begin{equation}
\hat{\Sigma}_{\mu\nu\rho}^{AB}=-4{\rm Tr}\,T^AT^B\, \Sigma_{\mu\nu\rho}\, , \label{eq:selfen3}\end{equation}
with $\Sigma_{\mu\nu\rho}$ given by (\ref{eq:3.23}) with $(f^2-\frac{1}{2}g^2)
\rightarrow f^2$.
Next we calculate the correction to the $U(1)$ vertex which makes it consistent
with gauge invariance.  Assume a correction of the form \begin{equation}
\Delta V_i^\mu(z,u,v)=\left(\frac{ig_1}{2}\right)\gamma^\mu(\alpha_i L+\beta_i R)\,\delta
^4(z-u)\delta^4(z-v) \, .\end{equation}
It's divergence is \begin{equation}
\partial_\mu^z\Delta V_i^\mu(z,u,v)=\frac{ig_1}{2}\,(\alpha_i R+\beta_i L)\,
[\delta^4(z-u)-\delta^4(z-v)] \,/\hspace{-2mm}
\partial_u\,\delta^4(u-v)\, . 
\label{eq:DV1}\end{equation}
Recall the $U(1)$ Ward identity \begin{equation}
\partial_\mu^z\Delta V_i^\mu(z,u,v)=-\frac{ig_1}{2}[\delta^4(z-u)-\delta^4(z-v)](Y_R^iL+Y_L^iR)
\Delta\Sigma^i(u-v)\, , \label{eq:WI1'}\end{equation}
where $\Delta\Sigma^i(u-v)$ is for our purposes \begin{equation}
\Delta\Sigma^i(u-v)=\Sigma_3^i(u-v)-\Sigma_1^i(u-v)=\frac{3f_d^2}{16(4\pi^2)}\left(
\frac{2}{Y_R^{(d)}}\delta_{i(d)}\,L-\frac{1}{Y_L^{(d)}}\, R\right)/\hspace{-2mm}
\partial_u\delta^4(u-v)\, . \label{eq:dS}\end{equation}
Comparing (\ref{eq:DV1}) with (\ref{eq:WI1'}) and (\ref{eq:dS}) determines
$\alpha$ and $\beta$, giving
\begin{equation}
\Delta V_i^\mu(z,u,v)=\frac{ig_1}{2}\left(\frac{3f_d^2}{16(4\pi^2)}\right)
\gamma^\mu\,[L-2\delta_{i(d)}\,R]\, \delta^4(z-u)\delta^4(z-v) \, . 
\end{equation}
Inserting this at the $U(1)$ vertex of the basic fermion triangle (and summing
over the $u$ and $d$ quarks) gives for the abnormal parity part 
\begin{equation}
\Delta V_{\mu\nu\rho}^{AB}=8{\rm Tr}\,T^AT^B\,\Sigma_{\mu\nu\rho}\, ,
\label{eq:vertcor1}\end{equation}
where again $\Sigma_{\mu\nu\rho}$ is given by (\ref{eq:3.23}) with $(f^2-
\frac{1}{2}g^2)\rightarrow f^2$.
Finally, the only nonplanar diagram contributes \begin{equation}
\hat{N}_{\mu\nu\rho}^{(z)\,AB}=-4{\rm Tr}\,T^AT^B\,N_{\mu\nu\rho}\, . 
\label{eq:nonplanar3}\end{equation}

Once again, summing the contributions from (\ref{eq:1-higgs31}), 
(\ref{eq:1-higgs33}),
(\ref{eq:selfen3}), (\ref{eq:vertcor1}) and (\ref{eq:nonplanar3}) gives zero 
total
contribution to the axial part of the $\langle J_1J_3J_3\rangle$
correlator, which is proportional to
\begin{equation}
\Big[ \underbrace{\left(8-4\cdot 3\right)\cdot
3}_{\stackrel{\scriptstyle{\rm vertex\,correction}}{{\scriptstyle{\rm + self\,energy}}}}-
\underbrace{4\cdot 1}_{{\scriptstyle{\rm vertex}}}-\underbrace{4\cdot({
\scriptstyle -}4)}_{{\scriptstyle{\rm nonplanar}}}\Big]\,\, Y_R^{(e)}\,
{\rm Tr}\,T^AT^B \,=0 . \end{equation}

\setcounter{section}{5}
\mysection{Supersymmetric Gauge Theories}

Techniques very similar to those of this paper were applied in a recent study
of the operator product algebra of conserved currents of SUSY gauge theories
\cite{SCFT}.  The required calculations were very briefly summarized in 
\cite{SCFT}, and we will discuss some aspects in more detail here.

$N=1$ SUSY gauge theories contain component fields $A_\mu^a(x)$ and $\lambda^a
(x)$, gluons and gluinos respectively, in the adjoint representation of a
gauge group $G$, and complex scalars $\phi^i$ and their spinor partners 
$\psi^i$ which transform in a representation of $G$ with Hermitian generators
$T^{a\,i}_j$.  There are gauge interactions with gauge coupling $g$ and a
cubic superpotential with complex coupling $Y_{ijk}$ which is totally
symmetric.  Using Euclidean Majorana spinors \cite{Nicolai}, the action is
\begin{eqnarray}
S &=&\int d^{4}x\left[ \frac{1}{4}F_{\mu \nu  
}^{~~2}+\frac{1}{2}\overline{\lambda}%
{D}\!\!\!\!\slash\lambda +\overline{D_{\mu }\phi }D_{\mu }\phi  
+\frac{1}{2}%
\overline{\psi}{D}\!\!\!\!\slash\psi \right.   \nonumber \\
&&+i\sqrt{2}g(\overline{\lambda}^{a}\,\overline{\phi}_{i}T_{~~j}^{ai}L\psi  
^{j}-\overline{\psi%
}_{i}\,R\,T_{~~j}^{ai}\phi ^{j}\lambda ^{a})  \nonumber \\
&&-\frac{1}{2}(\overline{\psi}^{i}\,L\,Y_{ijk}\phi ^{k}\psi  
^{j}+\overline{\psi}_{i}R\bar{Y%
}^{ijk}\overline{\phi}_{k}\psi _{j})  \nonumber \\
&&\left. +\frac{1}{2}g^{2}(\overline{\phi}_{i}T_{~~j}^{ai}\phi  
^{j})^{2}+\frac{1}{%
4}Y_{ijk}\overline{Y}^{ilm}\phi ^{j}\phi  
^{k}\overline{\phi}_{l}\overline{\phi}_{m}\right] \, .
\label{eq:lagra}
\end{eqnarray}

The theory has two classically conserved, but anomalous, axial currents,
the $R$-current and the Konishi current
\begin{eqnarray}
 &R_{\mu }(x)=\frac{1}{2}\overline{\lambda}\gamma _{\mu }\gamma  
_{5}\lambda -\frac{1%
}{6}\overline{\psi}\gamma _{\mu }\gamma _{5}\psi
+\frac{2}{3}\overline{\phi}\stackrel{\longleftrightarrow }{D}_{\mu }\phi&
\nonumber \\
 &K_{\mu }(x)=\frac{1}{2}\overline{\psi}_{i}\gamma _{\mu }\gamma _{5}\psi  
 ^{i}+\overline{%
\phi}_{i}\stackrel{\longleftrightarrow }{D}_{\mu }\phi ^{i} \, .&\end{eqnarray}
Actually $K_\mu$ is conserved classically only if $Y_{ijk}=0$.  The operator
prodect algebra of $R_\mu$ and $K_\mu$ contains two central charges $c$ and $c'$, and
the principal result of \cite{SCFT} was evidence for a universality property
of the interaction dependent radiative corrections to $c$ and $c'$.  For
$c'$ this
information was obtained from a study of two-loop contributions to the two
and three-point correlation functions of the currents, including the correlator
$\langle R_\mu(x)R_\nu(y)K_\rho(z)\rangle$.  Earlier work \cite{Jack} could 
be modified to obtain the required information about $c$.
Internal gauge and superpotential interactions can be treated separately
in two-loop order.  

In the gauge sector, both currents are conserved and
have no anomalous dimensions, and conformal invariance holds, so the
amplitude is again a constant multiple of the unique conformal pseudotensor
$A_{\mu\nu\rho}$ of (\ref{eq:2.9}).  The Feynman graphs of
$\langle R_\mu R_\nu K_\rho\rangle$ involve the gluon and Yukawa interactions 
of
(\ref{eq:lagra}) after setting $Y_{ijk}=0$, and they are the same graphs
considered in Sections 3-5 above with different numerical coefficients.  As
was the case in the standard model, there are Ward identities relating vertex 
functions of $R_\mu(x)$ and $K_\mu(x)$ to the same self-energy $\Sigma(u-v)$,
and an additional local term is required for one of the currents.  There
is some freedom in the assignment of local terms to the currents and 
self-energy, but the full correlator $\langle R_\mu R_\nu K_\rho\rangle$ is
independent of the choice made.  Using one convenient choice, and after
careful comparison of all graphs with those of the basic $U(1)$ model of
Section 3, it was found that the sum of all one and two-loop contributions is
\begin{equation}
\langle R_\mu(x) R_\nu(y) K_\rho(z)\rangle=\left[\frac{1}{9}{\rm dim}\,T+
\frac{g^2}{32\pi^2}\,{\rm Tr}\,T^aT^a\left(\frac{8}{3}+\frac{1}{3}-3\right)
\right]A_{\mu\nu\rho}(x,y,z) \, ,\label{eq:6.3} \end{equation}
where $A_{\mu\nu\rho}$ is given in (\ref{eq:2.9}).  The order $g^2$ two-loop
amplitude vanishes, with nonplanar, vertex and self-energy graphs 
contributing in the ratio $8:1:-9$, which is different from the ratios in
(\ref{eq:3.30}).  The net result is an Adler-Bardeen theorem for the
$\langle R_\mu R_\nu K_\rho\rangle$ correlator, since the sum of virtual
gluon graphs can again be shown to vanish by previous work \cite{BJ}.

The effect of the superpotential interactions was also considered in 
\cite{SCFT}.
However it was simpler to replace the current $K_\rho(z)$ by its scalar
superpartner $K(z)=\overline{\phi}(z)\phi(z)$, which is a scalar mass
operator of canonical dimension two, and study the correlator
$\langle R_\mu(x) R_\nu(y) K(z)\rangle$.  The operator $K(z)$ (as well as 
$K_\rho(z)$)
acquires an anomalous dimension of order $Y_{ijk}\overline{Y}^{ijk}$.  An anomalous
dimension is consistent with conformal symmetry, and the correlator can be 
shown to be conformal covariant through two-loop order.  Inversion,
conservation and scale properties can be used to fix its tensor form up to a
multiplicative constant \cite{SCFT}.  The graphs contributing to
$\langle R_\mu R_\nu K\rangle$ are typically subdivergent because they contain
subdiagrams with gauge invariant anomalous dimension.  Nevertheless in this
more complicated situation the conformal inversion technique could be combined 
with differential regularization \cite{Bible} to compute all contributing
Feynman diagrams.

All results for $c'$ obtained by the conformal methodology of this paper
were verified by an alternate method of calculation in which the four-point
correlation function $\langle R_\mu(x)R_\nu(y)R_\rho(z)R_\sigma(w)\rangle$
was studied in the relevant asymptotic region using regularization by
dimensional reduction in intermediate stages of the calculation.  The
explicit use of Ward identities to determine ambiguous local terms in
self-energy and vertex insertions was not required in this approach, so
agreement of the results of the two methods provides a check on this aspect
of the conformal approach.

\vspace{0.3in}
\noindent{\bf Acknowledgements}

\noindent We are grateful to K. Johnson for sharing his insights into the
space-time approach to quantum field theory, and to R. Jackiw and H. Osborn
for useful discussions and for their interest in this study.  D.Z.F. thanks
H. Cheng for a steady exchange of ideas over the course of this work.
The research of D.Z.F. was supported by NSF grant PHY-92-06867.  The research
of J.E. was supported by DOE cooperative research
agreement DE-FC02-94ER40818.

\renewcommand{\theequation}{A.\arabic{equation}}
\appendix \mysection{Appendix}

We discuss here the convolution integrals required in Section 3 to elucidate
the finite gauge mechanism and determine the local part of the self-energy,
and to calculate the two-loop nonplanar and vertex insertion diagrams for
the anomalous correlation function $\langle J_\mu(z)J_\nu(x)J_\rho(y)\rangle$.

We use the method \cite{Rosner} of Gegenbauer polynomials, which appear 
naturally because their generating function is the scalar propagator
\begin{eqnarray}
&{\displaystyle \frac{1}{(x-y)^2}\equiv \frac{1}{x^2}\,\sum_{n=0}^{\infty}\,\left(\frac{y}
{x}\right)^n\,C_n(\hat{x}\cdot\hat{y})\hspace{0.4in}|x|>|y|}& \nonumber \\
&{\displaystyle \hat{x}_\mu=\frac{x_\mu}{|x|}\hspace{0.3in} \hat{y}_\mu=\frac{y_\mu}{|y|}
\hspace{0.3in}\hat{x}\cdot \hat{y}=\cos\theta}& \label{eq:A1}\end{eqnarray}
\begin{equation}
C_n(\cos\theta)=\frac{\sin(n+1)\theta}{\sin\theta} \, .\label{eq:A2}
\end{equation}
The orthogonality relation obeyed by these polynomials is \begin{equation}
\int d\hat{x}\,C_n(\hat{x}\cdot\hat{y})\,C_m(\hat{x}\cdot\hat{z})=2\pi^2
\,\delta_{nm}\,\frac{C_n(\hat{y}\cdot\hat{z})}{n+1} \, .
\label{eq:A3} \end{equation}
where $d\hat{x}=\sin^2\theta\,\sin\phi\,d\theta\,d\phi\,d\xi$ is the angular
integration measure for the $\hat{x}$ variable.

We first list the integrals, and then comment briefly on their evaluation.
We use the notation $\Delta=x-y$.
\begin{eqnarray}
\int\frac{d^4v}{v^2(v-x)^2}&=&-\pi^2\,\ln\frac{x^2}{R^2} \label{eq:A4}\\
\int d^4v\,\frac{(v-x)_\rho}{v^2(v-x)^4}&=&-\pi^2\,\frac{x_\rho}{x^2} 
\label{eq:A5} \\
\int d^4v\,\frac{(v-x)_\rho\,(v-y)_\sigma}{(v-x)^4(v-y)^4}&=&\frac{\pi^2}{2
\Delta^2}\left(\delta_{\rho\sigma}-\frac{2\Delta_\rho\Delta_\sigma}{\Delta^2}
\right) \label{eq:A6} \\
\int d^4v\, \frac{\left(v_\rho v_\sigma-\frac{1}{4}\delta_{\rho\sigma}\,v^2
\right)}{v^4(v-x)^2}&=&\frac{\pi^2}{2x^2}\left(x_\rho x_\sigma-\frac{1}{4}x^2
\delta_{\rho\sigma}\right) \label{eq:A7} \\
\int d^4v\,\frac{\left((v-x)_\rho(v-x)_\sigma-\frac{1}{4}\delta_{\mu\sigma}
(v-x)^2\right)\,(v-y)_\lambda}
{(v-x)^4(v-y)^4}&=&\frac{-\pi^2}{4\Delta^2}\left(
\delta_{\rho\lambda}\Delta_\sigma+\delta_{\sigma\lambda}\Delta_\rho-2\frac{
\Delta_\rho\Delta_\sigma\Delta_\lambda}{\Delta^2}\right)\label{eq:A8} \\
\int d^4v\,\frac{\left(v_\rho v_\sigma-\frac{1}{4}\delta_{\rho\sigma}v^2\right)}{v^6(v-x)^2}
&=& \frac{\pi^2}{2x^4}\left(x_\rho x_\sigma-\frac{1}{4}x^2\,\delta_{\rho\sigma}
\right)\label{eq:A9} \end{eqnarray}
\begin{eqnarray}
\lefteqn{\int d^4v\,\frac{\left((v-x)_\rho(v-x)_\sigma-\frac{1}{4}\delta_
{\rho\sigma}\,(v-x)^2\right)\,(v-y)_\lambda}{(v-x)^6(v-y)^4}} \hspace{1in}
\nonumber \\
 & &=\frac{-\pi^2}{8}\left[\frac{2\delta_{
\lambda\rho}\Delta_\sigma+2\delta_{\lambda\sigma}\Delta_\rho+\delta_{\rho
\sigma}\Delta_\lambda}{\Delta^4}-\frac{8\Delta_\lambda\Delta_\rho\Delta_\sigma}
{\Delta^6}\right]
\label{eq:A10} \end{eqnarray}

To evaluate (\ref{eq:A4}) one applies (\ref{eq:A1}) to the factor $1/(v-x)^2$
including both regions $v<x$ and $v>x$, which have different dependence
on the radial variable $v$.  The quantity $R$ is a temporary cutoff which has
no effect on the integrals used in Section 3.  The result (\ref{eq:A5}) is
obtained by differentiation of (\ref{eq:A4}), and (\ref{eq:A6}) is obtained
by replacing $x\rightarrow x-y=\Delta$ in (\ref{eq:A5}), changing integration
variables to $v'=v+y$, and then differentiating with respect to $y$.
To evaluate (\ref{eq:A7}) or (\ref{eq:A9}) one takes the scalar product with
$x_\rho x_\sigma$, so that the integral contains the explicit Gegenbauer
polynomial $C_2(\hat{x}\cdot\hat{v})$.  One then applies (\ref{eq:A1}) to the
factor $1/(v-x)^2$ and uses orthogonality (\ref{eq:A3}).  Finally,
(\ref{eq:A8}) and (\ref{eq:A10}) are obtained from (\ref{eq:A7}) and
(\ref{eq:A9}) respectively by replacement $x\rightarrow x-y=\Delta$, shift
of integration variables and differentiation with respect to $y$.

\end{document}